\pdfoutput=1 %Added for uploading to ARXIV
\documentclass[pmlr]{jmlr}% new name PMLR (Proceedings of Machine Learning)

\usepackage{rotating}% for sideways figures and tables
\usepackage{longtable}% for long tables

\usepackage{booktabs}
\usepackage[load-configurations=version-1]{siunitx} % newer version

\usepackage{tikz}
\usepackage{multirow}
\usepackage{multicol}
\usepackage{graphicx}
\usepackage[normalem]{ulem}
\useunder{\uline}{\ul}{}
\usepackage{pgfplots}
\usepgfplotslibrary{groupplots}
\pgfplotsset{compat=1.14}

\theorembodyfont{\upshape}
\theoremheaderfont{\scshape}
\theorempostheader{:}
\theoremsep{\newline}

\jmlrvolume{85}
\jmlryear{2019}
\jmlrworkshop{Machine Learning for Healthcare}

\title[Diabetes Forecasting]{Diabetes Mellitus Forecasting Using Population Health Data in Ontario, Canada}

\author{\Name{Mathieu Ravaut} \Email{mathieu@layer6.ai} 
       \addr Layer 6 AI
       \AND
       \Name{Hamed Sadeghi} \Email{hamed@layer6.ai} 
       \addr Layer 6 AI
       \AND
       \Name{Kin Kwan Leung} \Email{kk@layer6.ai} 
       \addr Layer 6 AI
       \AND
       \Name{Maksims Volkovs} \Email{maks@layer6.ai} 
       \addr Layer 6 AI
       \AND
       \Name{Laura C. Rosella} \Email{laura.rosella@utoronto.ca}
       \addr Epidemiology Division, Dalla Lanna School of Public Health, University of Toronto
       }

\begin{document}
%%%%%%%%%%%%%%%%%%%%%%%%%%%%%%%%%%%%%%%%%%%%%%%%%
% Dataset Pruning
%%%%%%%%%%%%%%%%%%%%%%%%%%%%%%%%%%%%%%%%%%%%%%%%%
\newcommand{\tableDatasetPruning}
{
\begin{table}[!h]
\caption{Training details for Dataset Pruning. Top features and aggregated Shapley values are also displayed. The model is XGBoost-avg with $w=5$.}
\resizebox{\textwidth}{!}{
\begin{tabular}{|c|c|c|c|c|c|c|c|c|c|c|c|c|}
\hline
\multirow{2}{*}{Dataset(s)} & \multicolumn{2}{|c|}{AUC} & \multicolumn{2}{|c|}{Accuracy} & \multicolumn{2}{|c|}{Sensitivity} & \multicolumn{2}{|c|}{Specificity} & \multicolumn{2}{|c|}{PPV} & \multirow{2}{*}{Top Features}                                                                                                                                                         & \multirow{2}{*}{Values}                                                       \\
                            & train      & test      & train         & test        & train          & test          & train          & test          & train      & test      &                                                                                                                                                                                       &                                                                               \\ \hline
all                         & 82.2       & 79.9       & 74.2          & 71.9         & 79           & 72.2           & 68.9           & 71.5           & 74.4       & 74.1       &          -                                                                                                                                                                             & -                                                                              \\ \hline
fixed                       & 65       & 62.6       & 60.7          & 59.3         & 81.6           & 76.8           & 36.9              & 39.5           & 59.6       & 58.9       & \begin{tabular}[c]{@{}c@{}}Diagnosis Age\\ Deprivation Quintile\\ Nearest Public Health Unit\\ Birth Year\\ Country of Citizenship (2)\end{tabular}                                   & \begin{tabular}[c]{@{}c@{}}0.471\\ 0.043\\ 0.042\\ 0.036\\ 0.033\end{tabular} \\ \hline
chronic                     & 62.9         & 63.1       & 61.8            & 61.7           & 60.6           & 58.9           & 63.1           & 64.9           & 65.3       & 65.5       & \begin{tabular}[c]{@{}c@{}} ORAD Incidence\\ Asthma Prevalence\\CHF Prevalence\\ Asthma Incidence\\ COPD Prevalence\end{tabular}                                                      & \begin{tabular}[c]{@{}c@{}}0.453\\ 0.137\\ 0.046\\ 0.028\\ 0.017\end{tabular} \\ \hline
DAD                         & 57.3       & 56.7       & 55            & 54.8         & 42.5           & 39.8             & 69.2           & 71.8             & 61.2       & 61.5       & \begin{tabular}[c]{@{}c@{}}DAD Observation Count\\ source (i)\\ source (s)\\ Diagnostic Type IX\\ Admission Category (u)\end{tabular}                                                               & \begin{tabular}[c]{@{}c@{}}0.461\\ 0.024\\ 0.013\\ 0.006\\ 0.006\end{tabular} \\ \hline
ERCLAIM                     & 58.8       & 57.9       & 56.5          & 56.4         & 52.3           & 52.4           & 61.2           & 61           & 60.7       & 60.3       & \begin{tabular}[c]{@{}c@{}}ERCLAIM Observation Count\\ OHIP Location Claim (E)\\ Max OHIP Location Claim (E)\\ Physician Specialty (Family/General)\\ Max Total fee paid to Physician\end{tabular} & \begin{tabular}[c]{@{}c@{}}0.399\\ 0.066\\ 0.036\\ 0.024\\ 0.023\end{tabular} \\ \hline
NACRS                       & 58.8       & 58.6         & 56.9          & 56.9         & 56.5           & 55.7           & 57.4           & 58.3           & 60.3       & 60.2       & \begin{tabular}[c]{@{}c@{}}NACRS Observation Count\\ Blood Transfused (no) \\ Max Blood Transfused (no)\\ Main Problem (ICD10 IX)\\ Max Main Problem (ICD10 XVIII) \end{tabular}       & \begin{tabular}[c]{@{}c@{}}0.461\\ 0.030\\ 0.016\\ 0.012\\ 0.008\end{tabular} \\ \hline
ODB                         & 58.4       & 59.4       & 56.2          & 57.8         & 42.2           & 50.6             & 72.3           & 66.1             & 63.5         & 62.8       & \begin{tabular}[c]{@{}c@{}}ODB Observation Count\\ Quantity\\ Max Quantity\\ Long Term Care (no)\\ Max Long Term Care (no)\end{tabular}                                                 & \begin{tabular}[c]{@{}c@{}}0.396\\ 0.090\\ 0.062\\ 0.020\\ 0.011\end{tabular} \\ \hline
OHIP                        & 74.8       & 69.8       & 68.4          & 65.2         & 74.2           & 69.5           & 61.8           & 60.4           & 68.9       & 66.5       & \begin{tabular}[c]{@{}c@{}}OHIP Observation Count\\ (ICD9) Hypertension\\ Max (ICD9) Hypertension\\ (ICD9) Immunity\\ (ICD9) Diabetes\end{tabular}                                & \begin{tabular}[c]{@{}c@{}}0.559\\ 0.198\\ 0.116\\ 0.113\\ 0.092\end{tabular}   \\ \hline
OLIS                        & 66.2       & 72.6       & 60.4          & 68         & 36.4           & 60.2             & 87.9           & 76.8           & 77.4       & 74.6       & \begin{tabular}[c]{@{}c@{}}A1c\\ Max (LOINC) 4548-4\\ Max Range\\ Max (LOINC) 14771-0\\ Max (LOINC) 71875-9 \end{tabular}                                                              & \begin{tabular}[c]{@{}c@{}}0.66\\ 0.228\\ 0.214\\ 0.210\\ 0.193\end{tabular}  \\ \hline
\end{tabular}}
\end{table}
}

%%%%%%%%%%%%%%%%%%%%%%%%%%%%%%%%%%%%%%%%%%%%%%%%%
% XGBoost xvg Top Feature Contributions (Updated Mar25 3:06pm)
%%%%%%%%%%%%%%%%%%%%%%%%%%%%%%%%%%%%%%%%%%%%%%%%%
\newcommand{\tableXGBAvgTopFeatures}
{
\begin{table}[!h]
\centering
\caption{XGBoost-avg Top Feature Contributions}
\begin{tabular}{|c|c|}
\hline
Features & Aggregated Shapley Values \\ \hline
OLIS: A1c & 0.908 \\
OLIS: Range  &0.282 \\
OLIS: Max LOINC 4548-4 & 0.222 \\
OLIS: Max LOINC 14771-0 & 0.176 \\
OLIS: Max Range & 0.164 \\
OLIS: Max LOINC 71875-9 & 0.154 \\
Asthma:  Prevalence & 0.141 \\
OLIS: Max LOINC 14749-6 & 0.141 \\
OHIP: Hypertension Claim Code & 0.116 \\
OHIP: Immunity Claim Code & 0.096 \\
Nearest Public Health Unit & 0.076 \\
Age & 0.068 \\
OLIS: LOINC 14749-6 & 0.060 \\
Birth Year & 0.056 \\
\hline
\end{tabular}
\end{table}
}

%%%%%%%%%%%%%%%%%%%%%%%%%%%%%%%%%%%%%%%%%%%%%%%%%
% LR vg Top Feature Contributions (Updated Mar25 3:06pm)
%%%%%%%%%%%%%%%%%%%%%%%%%%%%%%%%%%%%%%%%%%%%%%%%%
\newcommand{\tableLRAvgTopFeatures}
{
\begin{table}[!h]
\centering
\caption{LR-avg Top Feature Contributions}
\begin{tabular}{|c|c|}
\hline
Features & Aggregated Shapley Values \\ \hline
OLIS: LOINC 14771-0 & 0.665 \\
OLIS: LOINC 14749-6 & 0.632 \\
OLIS: Max LOINC 4548-4 & 0.321 \\
LHIN Central & 0.302 \\
Landing date & 0.277 \\
LHIN Central East & 0.258 \\
LHIN Champlain & 0.239 \\
LHIN Hamilton Niagara & 0.23 \\
LHIN Missisauga Halton & 0.214 \\
OLIS: Max LOINC 71875-9 & 0.209 \\
LHIN Central West & 0.164 \\
LHIN Toronto Central & 0.156 \\
LHIN South West & 0.154 \\
LHIN North East & 0.12 \\
OHIP: Max Immunity Claim Code & 0.096 \\
\hline
\end{tabular}
\end{table}
}

%%%%%%%%%%%%%%%%%%%%%%%%%%%%%%%%%%%%%%%%%%%%%%%%%
% Highway avg Top Feature Contributions (Updated Mar25 3:06pm)
%%%%%%%%%%%%%%%%%%%%%%%%%%%%%%%%%%%%%%%%%%%%%%%%%
\newcommand{\tableHighwayAvgTopFeatures}
{
\begin{table}[!h]
\centering
\caption{Highway-avg Top Feature Contributions}
\begin{tabular}{|c|c|}
\hline
Features & Aggregated Shapley Values \\ \hline
OLIS: Max Range & 1.964 \\
OLIS: Range & 1.742 \\
OLIS: LOINC 14771-0 & 0.407 \\
OLIS: LOINC 14749-6 & 0.375 \\
Longitude & 0.351 \\
OLIS: Max LOINC 14749-6 & 0.340 \\
OLIS: Max LOINC 14771-0 & 0.335 \\
Age & 0.310 \\
Birth Year & 0.310 \\
Rurality (No) & 0.226 \\
OLIS: Max LOINC 4548-4 & 0.186 \\
OHIP: Max Immunity Claim Code & 0.143 \\
OHIP: Max Hypertension Claim Code & 0.118 \\
OLIS: Max LOINC 71875-9 & 0.113 \\
OHIP: Max Location (o) & 0.1\\
\hline
\end{tabular}
\end{table}
}

%%%%%%%%%%%%%%%%%%%%%%%%%%%%%%%%%%%%%%%%%%%%%%%%%
% XGBoost Parameters
%%%%%%%%%%%%%%%%%%%%%%%%%%%%%%%%%%%%%%%%%%%%%%%%%
\newcommand{\tableXGBParameters}
{
\begin{table}[!h]
\centering
\caption{XGBoost Parameters}
\begin{tabular}{|c|c||c|c|}
\hline
parameters & values & parameters & values\\ \hline \hline
Max depth & 30 & min child weight (\% of training data) & 0.001\\
Learning rate & 0.01 & alpha (L2 reg.) &  0.3\\ 
Subsample & 1 & lambda (L1 reg.) & 0.5 \\
Colsample by tree & 0.7 & gamma (\#tree reg.) & 0.1 \\
Colsample by level & 0.7 & Num Rounds & 500\\
\hline
\end{tabular}
\end{table}
}

%%%%%%%%%%%%%%%%%%%%%%%%%%%%%%%%%%%%%%%%%%%%%%%%%
% LR Parameters
%%%%%%%%%%%%%%%%%%%%%%%%%%%%%%%%%%%%%%%%%%%%%%%%%
\newcommand{\tableLRParameters}
{
\begin{table}[!h]
\centering
\caption{Logistic Regression Parameters in scikit-learn}
\begin{tabular}{|c|c||c|c|}
\hline
parameters & values & parameters & values\\ \hline \hline
penalty & L1 & class\_weight & None\\
dual & false & random\_state & None\\ 
tol & 0.0001 & solver & liblinear \\
C & 1 & max\_iter & 100 \\
fit\_intercept & true & multi\_class & warn\\
intercept\_scaling & 1 & warm\_start & false\\
\hline
\end{tabular}
\end{table}
}

%%%%%%%%%%%%%%%%%%%%%%%%%%%%%%%%%%%%%%%%%%%%%%%%%
% Highway Parameters
%%%%%%%%%%%%%%%%%%%%%%%%%%%%%%%%%%%%%%%%%%%%%%%%%
\newcommand{\tableHighwayParameters}
{
\begin{table}[!h]
\centering
\caption{Highway Parameters}
\begin{tabular}{|c|c||c|c|}
\hline
parameters & values & parameters & values\\ \hline \hline
Learning rate & $10^{-5}$ & Batch size &  128\\
Dropout & 0.5 & Early Stopping & true\\
\hline
\end{tabular}
\end{table}
}

%%%%%%%%%%%%%%%%%%%%%%%%%%%%%%%%%%%%%%%%%%%%%%%%%
% CNN-LSTM Parameters
%%%%%%%%%%%%%%%%%%%%%%%%%%%%%%%%%%%%%%%%%%%%%%%%%
\newcommand{\tableCNNParameters}
{
\begin{table}[!h]
\centering
\caption{CNN-LSTM Parameters}
\begin{tabular}{|c|c||c|c|}
\hline
parameters & values & parameters & values\\ \hline \hline
Learning rate & $10^{-5}$ & Batch size &  128\\
Dropout & 0.5 & Early Stopping & true\\
\hline
\end{tabular}
\end{table}
}

%%%%%%%%%%%%%%%%%%%%%%%%%%%%%%%%%%%%%%%%%%%%%%%%%
% LSTM-Seq2Seq Parameters
%%%%%%%%%%%%%%%%%%%%%%%%%%%%%%%%%%%%%%%%%%%%%%%%%
\newcommand{\tableLSTMParameters}
{
\begin{table}[!h]
\centering
\caption{LSTM-Seq2Seq Parameters}
\begin{tabular}{|c|c||c|c|}
\hline
parameters & values & parameters & values\\ \hline \hline
Learning rate & $10^{-5}$ & Batch size &  128\\
Dropout & 0.2 & Early Stopping & true\\
\hline
\end{tabular}
\end{table}
}

%%%%%%%%%%%%%%%%%%%%%%%%%%%%%%%%%%%%%%%%%%%%%%%%%
% LR Results
%%%%%%%%%%%%%%%%%%%%%%%%%%%%%%%%%%%%%%%%%%%%%%%%%
\newcommand{\tableLRResults}
{
\begin{table}[!h]
\centering
\caption{Logistic Regression Results for $b=1$}
\begin{tabular}{|c|c|c|c|c|c|c|}
\hline
Type   & Window & AUC  & Accuracy & Sensitivity & Specificity & PPV  \\ \hline
avg    & 1      & 74.2 & 67.7     & 68.6        & 66.8        & 70   \\
avg    & 3      & 77   & 70       & 68.7        & 71.5        & 73.2 \\
avg    & 5      & 77.6 & 70.6     & 67.7        & 74          & 74.6 \\
avg    & 10     & 77.7 & 70.8     & 67.2        & 74.8        & 75.1 \\
concat & 3      & 77.1 & 70       & 68.4        & 71.8        & 73.3 \\
concat & 5      & 77.8 & 71       & 68          & 74.3        & 75   \\
concat & 10     & 78   & 71.1     & 68.9        & 73.4        & 74.6 \\
\hline
\end{tabular}
\end{table}
}

%%%%%%%%%%%%%%%%%%%%%%%%%%%%%%%%%%%%%%%%%%%%%%%%%
% XGB Results
%%%%%%%%%%%%%%%%%%%%%%%%%%%%%%%%%%%%%%%%%%%%%%%%%
\newcommand{\tableXGBResults}
{
\begin{table}[!h]
\centering
\caption{XGBoost Results for $b=1$}
\begin{tabular}{|c|c|c|c|c|c|c|}
\hline
Type   & Window & AUC  & Accuracy & Sensitivity & Specificity & PPV  \\ \hline
avg    & 1      & 76.4 & 68.8     & 70.5        & 66.9        & 70.6 \\
avg    & 3      & 79.3 & 71.3     & 72.2        & 70.4        & 73.4 \\
avg    & 5      & 79.9 & 71.9     & 72.2        & 71.5        & 74.1 \\
avg    & 10     & 79.7 & 71.6     & 72.9        & 70.2        & 73.4 \\
concat & 3      & 79.6 & 71.6     & 70.4        & 73          & 74.7 \\
concat & 5      & 80.3 & 72.3     & 70.5        & 74.4        & 75.7 \\
concat & 10     & 79.7 & 71.6     & 72.9        & 70.2        & 73.4 \\
\hline
\end{tabular}
\end{table}
}

%%%%%%%%%%%%%%%%%%%%%%%%%%%%%%%%%%%%%%%%%%%%%%%%%
% Highway Results
%%%%%%%%%%%%%%%%%%%%%%%%%%%%%%%%%%%%%%%%%%%%%%%%%
\newcommand{\tableHighwayResults}
{
\begin{table}[!h]
\centering
\caption{Highway Results for $b=1$}
\begin{tabular}{|c|c|c|c|c|c|c|}
\hline
Type   & Window & AUC  & Accuracy & Sensitivity & Specificity & PPV  \\ \hline
avg    & 1      & 74.4 & 67.9     & 74.4        & 60.5        & 68.1 \\
avg    & 3      & 77.5 & 70.5     & 72.6        & 68.1        & 72   \\
avg    & 5      & 78.2 & 71.2     & 72.5        & 69.6        & 73   \\
avg    & 10     & 78.6 & 71.7     & 70.3        & 73.2        & 74.8 \\
concat & 3      & 76.9 & 69.9     & 74          & 65.3        & 70.7 \\
concat & 5      & 77.2 & 69.9     & 66.8        & 73.4        & 74   \\
concat & 10     & 76.7 & 70.3     & 75.4        & 64.6        & 70.6 \\
\hline
\end{tabular}
\end{table}
}

%%%%%%%%%%%%%%%%%%%%%%%%%%%%%%%%%%%%%%%%%%%%%%%%%
% CNN-LSTM Results
%%%%%%%%%%%%%%%%%%%%%%%%%%%%%%%%%%%%%%%%%%%%%%%%%
\newcommand{\tableCNNResults}
{
\begin{table}[!h]
\centering
\caption{CNN-LSTM Results for $b=1$}
\begin{tabular}{|c|c|c|c|c|c|}
\hline
Window & AUC  & Accuracy & Sensitivity & Specificity & PPV  \\ \hline
3      & 76.6 & 69.3     & 72.6        & 65.6        & 70.5 \\
5      & 77.5 & 70.2     & 72.1        & 68.1        & 71.9 \\
10     & 78   & 71       & 77.1        & 64.1        & 70.8 \\
\hline
\end{tabular}
\end{table}
}

%%%%%%%%%%%%%%%%%%%%%%%%%%%%%%%%%%%%%%%%%%%%%%%%%
% Lstm Seq2Seq Results
%%%%%%%%%%%%%%%%%%%%%%%%%%%%%%%%%%%%%%%%%%%%%%%%%
\newcommand{\tableLSTMResults}
{
\begin{table}[!h]
\centering
\caption{LSTM-Seq2Seq Results for $b=1$}
\begin{tabular}{|c|c|c|c|c|c|}
\hline
Window & AUC  & Accuracy & Sensitivity & Specificity & PPV  \\ \hline
3      & 76.3 & 68.9     & 64.5        & 74        & 73.6 \\
5      & 77.4 & 70.2     & 68.7        & 71.9       & 73.4 \\
10     & 78.4   & 71.3       & 72.3        & 70.1        & 73.2 \\
\hline
\end{tabular}
\end{table}
}

%%%%%%%%%%%%%%%%%%%%%%%%%%%%%%%%%%%%%%%%%%%%%%%%%
% ICD10 code map
%%%%%%%%%%%%%%%%%%%%%%%%%%%%%%%%%%%%%%%%%%%%%%%%%
\newcommand{\tableICDTenCodeMap}
{
\begin{table}[!h]
\centering
\caption{ICD10 code map}
\begin{tabular}{|c|c|}
\hline
ICD10 code & Category \\ \hline
A, B & I \\
C & II \\
D00 - D48 & II \\
D51 - D99 & III \\
D49 - D50 & unknown \\
E10 - E14 & diabetes \\
E00 - E09, E15 - E99 & IV \\
F & V \\
G & VI \\
H00 - H59 & VII \\
H60 - H95 & VIII \\
I & IX \\
J & X \\
K & XI \\
L & XII \\
M & XIII \\
N & XIV \\
O & XV \\
P & XVI \\
Q & XVII \\
R & XVIII \\
S, T & XIX \\
V,W,X,Y & XX \\
Z & XXI \\
U & XXII \\
Others & unknown \\ \hline
\end{tabular}
\end{table}
}

%%%%%%%%%%%%%%%%%%%%%%%%%%%%%%%%%%%%%%%%%%%%%%%%%
% ICD9 code map
%%%%%%%%%%%%%%%%%%%%%%%%%%%%%%%%%%%%%%%%%%%%%%%%%
\newcommand{\tableICDNineCodeMap}
{
\begin{table}[!h]
\centering
\caption{ICD9 code map}
\begin{tabular}{|c|c|}
\hline
ICD9 code & Category \\ \hline
001 - 139 & infectious \\
140 - 239 & neoplasms \\
250 & diabetes \\
244.9 & hypothynoidism \\
240 - 249, 251 - 277, 279 (exclude 244.9) & immunity \\
278 & obesity \\
285.9 & anemia \\
280 - 289 (exclude 285.9) & blood \\
290 - 319 & mental \\
327.23 & sleep apnea \\
320 - 389 (exclude 327.23) & nervous \\
401 & hypertension \\
390 - 400, 402 - 459 & circulatory \\
460 - 519 & respiratory \\
571.8 & chronic liver disease \\
520 - 579 (exclude 571.8) & digestive \\
580 - 629 & genitourinary \\
630 - 679 & pregnancy \\
680 - 709 & skin \\
710 - 739 & musculoskeletal \\
740 - 759 & congenital \\
760 - 779 & perinatal \\
780.53 & hypersomnia \\
790.6 & blood chemistry \\
790.21 & fasting glucose \\
790.29 & abnormal glucose \\
Starts with E & E \\
Starts with V & V \\
Others & ill-defined \\
\hline
\end{tabular}
\end{table}
}

%%%%%%%%%%%%%%%%%%%%%%%%%%%%%%%%%%%%%%%%%%%%%%%%%
% Country demographic XGB-avg Top features
%%%%%%%%%%%%%%%%%%%%%%%%%%%%%%%%%%%%%%%%%%%%%%%%%
\newcommand{\tableCountryXGBTopFeatures}
{
\begin{table}[!h]
\centering
\caption{Top 5 features on country of birth demographics for XGBoost-avg with $w=5$.}
\resizebox{\textwidth}{!}{
\begin{tabular}{|c|c|c|c|c|c|c|}
\hline
                        & Canada  & India    & China & Philippines   & Pakistan & Sri Lanka    \\ \hline
A1c                     & 0.874  & 0.627 & 0.891 & 0.658 & 0.632 & 0.759   \\
OLIS: Range             & 0.311  & 0.261 & 0.241 & 0.331 & 0.283 & 0.239   \\
OLIS: LOINC 4548-4 Max  & 0.189  & 0.194 & -     & 0.274 & 0.189 & 0.253 \\
OLIS: LOINC 14771-0 Max & 0.159  & 0.191 & 0.176 & -     & 0.166 & 0.242    \\
OLIS: LOINC 71875-9 Max &       -& 0.178 & 0.187 & 0.231 & 0.137 & 0.235  \\
OLIS: Range Max         & 0.18   & -     & -     & 0.198 & 0.159 & -      \\
Landing Date            &       -& -     & \bf{0.196} & -     & -     &  -   \\ \hline
\end{tabular}}
\label{tab:contrib-country}
\end{table}
}

\newcommand{\tableCountryModels}
{
\begin{table}[htbp]
\centering
\caption{Model performance on country of birth demographics. Here $w=5$ and $b=1$.}
\resizebox{\textwidth}{!}{
\begin{tabular}{|c|c|c|c|c|c|c|}
\hline
Country of Origin        & \#Test Individuals     & AUC  & Accuracy & Sensitivity & Specificity & PPV  \\ 
\hline
Canada & 18144 & 79   & 71       & 69.9        & 72.3        & 72.7 \\  
\hline
India    & 519   & 81   & 73.6     & 93.5        & 41.1        & 72.1 \\  
\hline
China    & 374   & 84.3 & 75.3     & 78          & 72.5        & 73.8 \\
\hline
Philippines    & 353   & 74.8 & 77.5     & 88.7        & 43.3        & 82.7 \\  
\hline
Pakistan    & 233   & 72.8 & 76       & 90.1        & 38.6        & 78.8 \\  
\hline
Sri Lanka    & 201   & 65   & 78.7     & 94          & 26.5        & 81.3 \\  
\hline
\end{tabular}}
\label{tab:model-country}
\end{table}
}

%%%%%%%%%%%%%%%%%%%%%%%%%%%%%%%%%%%%%%%%%%%%%%%%%
% Age demographic XGB-avg Top Features (Updated Mar25 3:17)
%%%%%%%%%%%%%%%%%%%%%%%%%%%%%%%%%%%%%%%%%%%%%%%%%
\newcommand{\tableAgeXGBTopFeatures}
{
\begin{table}[!h]
\centering
\caption{Top 5 features on age demographics for XGBoost-avg with $w=5$. }
  \resizebox{\textwidth}{!}{
\begin{tabular}{|c|c|c|c|c|c|c|c|c|}
\hline
                              & 20-29 & 30-39 & 40-49 & 50-59 & 60-69 & 70-79 & 80-89 & 90+   \\ \hline
A1c                           & 0.866 & 1.217 & 0.925 & 0.893 & 0.791 & 0.753 & 0.739 & 0.871 \\
OLIS: Range                   & 0.171 & 0.186 & 0.248 & 0.301 & 0.320 & 0.340 & 0.341 & 0.270 \\
OLIS: LOINC 4548-4 Max        & 0.164 & 0.153 & -     & 0.240 & 0.266 & 0.289 & 0.309 & 0.291 \\
OLIS: LOINC 14771-0 Max       &   -   & -     & 0.156 & 0.183 & 0.191 & -     & 0.205 & -     \\
OLIS: LOINC 14749-6 Max       &   -   & -     & -     & -     & -     & -     & 0.203 & 0.162 \\
OLIS: LOINC 71875-9 Max       &   -   & -     & -     & -     & -     & 0.189 & -     & -     \\
Current Age                   & \bf{0.168} & -     & -     & -     & -     & -     & -     & - \\
OLIS: Max Range               & -     & -     & 0.148 & 0.177 & 0.189 & 0.204 & -     & 0.161   \\
Nearest Public Health Unit    & -     & 0.134 & -     & -     & -     & -     & -     & -     \\
Asthma Prevalence             & \bf{0.155} & \bf{0.162} & \bf{0.161}  & -     & -     & -     & -     & -     \\ \hline
\end{tabular}}
\label{tab:contrib-age}
\end{table}
}

\newcommand{\tableAgeModels}
{
\begin{table}[htbp]
\centering
\caption{Model performance on age demographics. Here $w=5$ and $b=1$.}
\begin{tabular}{|c|c|c|c|c|c|c|}
\hline
Age                    & \#Test Individuals   & AUC  & Accuracy & Sensitivity & Specificity & PPV  \\ \hline
20-29 & 662  & 74   & 67.7     & 56.6        & 78.5        & 72.1 \\ 
\hline
30-39 & 1808 & 76.2 & 68.9     & 61.3        & 76.4        & 72.1 \\  
\hline
40-49 & 3251 & 81.1 & 72.9     & 69.5        & 76.4        & 74.9 \\  
\hline
50-59 & 5426 & 81   & 72.8     & 72.5        & 73.1        & 74.6 \\ 
\hline
60-69 & 4673 & 78.9 & 70.6     & 72.9        & 67.9        & 72.3 \\ \hline
70-79 & 2735 & 78.3 & 71.4     & 78          & 63.1        & 72.9 \\  
\hline
80-89 & 1016 & 76   & 70.7     & 79.9        & 58.9        & 71.3  \\
\hline
90+   & 248  & 76.6 & 70.1     & 75.9        & 63.7        & 67.8  \\ \hline
\end{tabular}
\label{tab:model-age}
\end{table}
}

%%%%%%%%%%%%%%%%%%%%%%%%%%%%%%%%%%%%%%%%%%%%%%%%%
% Gender demographic XGB-avg Top Features
%%%%%%%%%%%%%%%%%%%%%%%%%%%%%%%%%%%%%%%%%%%%%%%%%
\newcommand{\tableGenderXGBTopFeatures}
{
\begin{table}[!h]
\centering
\caption{Top 5 features on gender demographics for XGBoost-avg with $w=5$.}
\begin{tabular}{|c|c|c|}
\hline
Feature                       & male  & female \\ \hline
A1c                           & 0.982 & 0.888  \\
OLIS: Range                   & 0.247 & 0.313  \\
OLIS: Max LOINC 4548-4        & 0.204 & 0.281  \\
OLIS: Max LOINC 14771-0       & 0.167 &   -    \\
OLIS: Max Range               & 0.154 & 0.182  \\ 
OLIS: Max LOINC 14749-6       & -     & 0.159  \\ \hline
\end{tabular}
\label{tab:contrib-gender}
\end{table}
}

%%%%%%%%%%%%%%%%%%%%%%%%%%%%%%%%%%%%%%%%%%%%%%%%%
% Gender demographic Models
%%%%%%%%%%%%%%%%%%%%%%%%%%%%%%%%%%%%%%%%%%%%%%%%%
% \newcommand{\tableGenderModels}
% {
% \begin{table}[htbp]
% \centering
% \caption{Model performance on gender demographics. Here $w=5$ and $b=1$.}
% \resizebox{\textwidth}{!}{
% \begin{tabular}{|c|c|c|c|c|c|c|c|}
% \hline
% Gender                  & \#Test Individuals        & Model       & AUC  & Accuracy & Sensitivity & Specificity & PPV  \\ \hline
% \multirow{3}{*}{Male}   & \multirow{3}{*}{10914} & LR-avg      & 76.6 & 69.8     & 68.8        & 71          & 72.9 \\
%                         &                        & XGBoost-avg & 79.4 & 71.8     & 72.1        & 71.5        & 74.1 \\
%                         &                        & Highway-avg & 77.1 & 70.3     & 75.7        & 64.2        & 70.6 \\ \hline
% \multirow{3}{*}{Female} & \multirow{3}{*}{9652}  & LR-avg      & 76.4 & 69.9     & 66.9        & 73          & 73   \\
%                         &                        & XGBoost-avg & 79.1 & 71       & 71.8        & 70.2        & 72.4 \\
%                         &                        & Highway-avg & 77.1 & 70.4     & 71.8        & 68.8        & 71.4 \\ \hline
% \end{tabular}}
% \label{tab:model-gender}
% \end{table}
% }
\newcommand{\tableGenderModels}
{
\begin{table}[htbp]
\centering
\caption{Model performance on gender demographics. Here $w=5$ and $b=1$.}
\begin{tabular}{|c|c|c|c|c|c|c|}
\hline
Gender                  & \#Test Individuals        & AUC  & Accuracy & Sensitivity & Specificity & PPV \\  
\hline
Male   & 10565 & 79.4 & 71.8     & 72.1        & 71.5        & 74.1 \\ 
\hline
Female & 9259  & 79.1 & 71       & 71.8        & 70.2        & 72.4 \\
\hline
\end{tabular}
\label{tab:model-gender}
\end{table}
}

%%%%%%%%%%%%%%%%%%%%%%%%%%%%%%%%%%%%%%%%%%%%%%%%%
% All demogrphics
%%%%%%%%%%%%%%%%%%%%%%%%%%%%%%%%%%%%%%%%%%%%%%%%%
\newcommand{\tableAllDemographics}
{
\begin{table}[htbp]
  \centering 
  \caption{XGBoost-avg performance on different demographics for $w=5$ and $b=1$} 
  \resizebox{\textwidth}{!}{
  \begin{tabular}{cc}
  \multirow{3}{*}{
    \subtable[Age][c]{
  \begin{tabular}{|c|c|c|c|c|c|}\hline
Age                    & \#Test Individuals   & AUC  \\ \hline
20-29 & 662  & 73.9  \\ \hline
30-39 & 1808 & 75.8 \\ \hline
40-49 & 3251 & 81.6 \\ \hline
50-59 & 5426 & 81.6   \\ \hline
60-69 & 4673 & 79.6 \\ \hline
70-79 & 2735 & 79.1 \\ \hline
80-89 & 1016 & 76.4   \\ \hline
90+   & 248  & 76.8 \\ \hline
  \end{tabular}
  \label{tab:age} 
  } } & 
  \subtable[Gender]{
  \begin{tabular}{|c|c|c|}\hline
Gender & \#Test Individuals & AUC  \\ \hline
Male   & 10565 & 80 \\ \hline
Female & 9259  & 79.7 \\ \hline
  \end{tabular}
  \label{tab:gender} 
  } \\ \\ &
   \subtable[Country of origin]{
  \begin{tabular}{|c|c|c|}
\hline
Country of Origin        & \#Test Individuals     & AUC  \\ \hline
Canada & 18144 & 79.6 \\ \hline
India    & 519   & 81.9 \\ \hline
China    & 374   & 84.9 \\ \hline
Philippines    & 353   & 75.4 \\ \hline
Pakistan    & 233   & 75 \\ \hline
Sri Lanka    & 201   & 67.5   \\ \hline
  \end{tabular}
  \label{tab:country} 
  }
  \end{tabular}}
  \label{tab:demographic}
\end{table}
}

%%%%%%%%%%%%%%%%%%%%%%%%%%%%%%%%%%%%%%%%%%%%%%%%%
% Unbalanced demographic Models
%%%%%%%%%%%%%%%%%%%%%%%%%%%%%%%%%%%%%%%%%%%%%%%%%
\newcommand{\tableUnbalancedModels}
{
\begin{table}[htbp]
  \centering 
  \caption{Model performance on skewed test set, with buffer b = $1$ year. The models are optimal when considering the window size.} 
  \begin{tabular}{|c|c|c|c|c|c|}\hline
    Model & AUC & Accuracy & Sensitivity & Specificity & PPV \\ \hline
    LR-avg & 75.5 & \bf{73.1} & 65.9& \bf{73.7} & 17.9\\
    XGBoost-avg & \bf{81.6} & 71 & \bf{77.8} & 70.4 & \bf{18.1}\\
    Highway-avg & 77.8 & 72.6 & 70.5 & 72.8 & \bf{18.2} \\
    CNN-LSTM & 74.7 & 65 & 71.1 & 64.5 & 14.6 \\
    LSTM-Seq2Seq & 78.3 & 70.1 & 68.8 & 70.2 & 16.4 \\
    \hline
  \end{tabular}
  \label{tab:unbalanced} 
\end{table}
}

\newcommand{\tableAllFeatureContrib}
{
\begin{table}[htbp]
  \centering 
  \caption{Feature Contributions on \emph{test set}. Here $w=5$ and $b=1$. \newline \scriptsize{OLIS Range=Measurement (of any observation code) falls in range; Blood HbA1c=LOINC 4548-4; Fasting Glucose (in Serum/Plasma)=LOINC 14771.0; Glucose (in Serum/Plasma)=LOINC 14749-6; A1c Mass Fraction=LOINC 71875-9 \newline Max corresponds to the maximum within the year;\newline LHIN=Local Health Integration Networks} } 
  \resizebox{\textwidth}{!}{
  \begin{tabular}{ccc}
    \subtable[LR-avg]{
    \begin{tabular}{|c|c|}\hline
    Feature & Contrib. \\ \hline
    Fasting Glucose & 0.665 \\
    Glucose& 0.632 \\
    Max Blood HbA1c & 0.321 \\ 
    LHIN Central & 0.302 \\
    Landing date & 0.277 \\
    LHIN Central East & 0.258 \\
    LHIN Champlain & 0.239 \\
    LHIN Hamilton Niagara & 0.23 \\
    LHIN Mississauga Halton & 0.214 \\
    Max A1c Mass Fraction & 0.209 \\
    \hline
  \end{tabular}
  \label{tab:contrib-lr}
  }&
  \subtable[XGBoost-avg]{
   \begin{tabular}{|c|c|}\hline
    Feature & Contrib.\\ \hline
    A1c &0.908\\
    OLIS: Range  &0.282\\
    Max Blood HbA1c &0.222\\
    Fasting Glucose&0.176 \\
    OLIS: Max Range &0.164 \\
    Max A1c Mass Fraction & 0.154 \\
    Asthma:  Prevalence & 0.141 \\
    Max Glucose & 0.141 \\
    Hypertension OHIP Claim & 0.116 \\
    Immunity OHIP Claim & 0.096 \\
    \hline
  \end{tabular}
  \label{tab:contrib-xgb} 
  }&
  \subtable[Highway-avg][t]{
   \begin{tabular}{|c|c|}\hline
    Feature & Contrib.\\ \hline
    OLIS: Max Range & 1.964\\
    OLIS: Range  &1.742\\
    Fasting Glucose &0.407 \\
    Glucose & 0.375 \\
    Longitude & 0.351 \\
    Max Glucose & 0.340 \\
    Max Fasting Glucose & 0.335 \\
    Age & 0.310 \\
    Birth Year & 0.310 \\
    Rurality (No) & 0.226 \\
    \hline
  \end{tabular}
  \label{tab:contrib-highway} 
  }
  \end{tabular}
  }
\end{table}
}

%%%%%%%%%%%%%%%%%%%%%%%%%%%%%%%%%%%%%%%%%%%%%%%%%
% Window Size for all models
%%%%%%%%%%%%%%%%%%%%%%%%%%%%%%%%%%%%%%%%%%%%%%%%%
\newcommand{\tableWindowSize}
{
\begin{table}[htbp]
  \centering 
  \caption{Model results (AUC on test set) - Observation window sizes, b = 1 year} 
    \begin{tabular}{|c|c|c|c|c|}
    \hline
    \multirow{2}{*}{Model} & \multicolumn{4}{c|}{Observation Window size ($w$)} \\ \cline{2-5} 
                           & 1                       & 3      & 5     & 10    \\ \hline
    LR - avg               & \multirow{2}{*}{73.2}   & 75.9   & 76.5  & 76.5  \\ \cline{1-1} \cline{3-5} 
    LR - concat            &                         & 76     & 76.8  & 76.9  \\ \hline
    XGBoost - avg          & \multirow{2}{*}{75.4}   & 78.4   & 79.3  & 79.1  \\ \cline{1-1} \cline{3-5} 
    XGBoost - concat       &                         & 78.6   & 79.4  & 79.4  \\ \hline
    Highway - avg          & \multirow{2}{*}{73.4}   & 76.6   & 77.1  & 77.4  \\ \cline{1-1} \cline{3-5} 
    Highway - concat       &                         & 75.4   & 75.8  & 75    \\ \hline
    \end{tabular}
  \label{tab:result-window} 
\end{table}
}

%%%%%%%%%%%%%%%%%%%%%%%%%%%%%%%%%%%%%%%%%%%%%%%%%
% Results for all models
%%%%%%%%%%%%%%%%%%%%%%%%%%%%%%%%%%%%%%%%%%%%%%%%%
\newcommand{\tableResults}
{
\begin{table}[htbp]
  \centering 
  \caption{Model performance on test set, for a buffer b = 1 year.} 
  \begin{tabular}{|c|c|c|c|c|c|}\hline
    Model & AUC & Accuracy & Sensitivity & Specificity & PPV \\ \hline
    LR - avg & 77.7 & 70.8 & 67.2 & \bf{74.8} & 75.1 \\
    LR - concat & 78 & 71.1 & 68.9 & 73.4 & 74.6 \\
    XGBoost - avg & \bf{79.9} & \bf{71.9} & 72.2 & 71.5 & 74.1 \\
    XGBoost - concat & \bf{80.3} & \bf{72.3} & 70.5 & \bf{74.4} & \bf{75.7} \\
    Highway - avg & 78.6 & 71.7 & 70.3 & 73.2 & 74.8 \\
    Highway - concat & 77.2 & 69.9 & 66.8 & 73.4 & 74 \\
    CNN-LSTM & 78 & 71 & \bf{77.1} & 64.1 & 70.8 \\
    LSTM-Seq2Seq & 78.4 & 71.3 & 72.3 & 70.1 & 73.2 \\
    \hline
    weighted average & 81.1 & 72.8 & 72.2 & 73.5 & 75.5 \\
    \hline
  \end{tabular}
  \label{tab:result} 
\end{table}
}

%%%%%%%%%%%%%%%%%%%%%%%%%%%%%%%%%%%%%%%%%%%%%%%%%
% Train Test Split
%%%%%%%%%%%%%%%%%%%%%%%%%%%%%%%%%%%%%%%%%%%%%%%%%
\newcommand{\tableTrainTestSplit}
{
\begin{table}[htbp]
  \centering 
  \caption{Train-Test Split Statistics} 
  \begin{tabular}{|c|c|c|c|}\hline
    Data & \#Positives & \#Negatives & \#Total \\ 
    \hline
    Train & 67,499 (53.3\%) & 59,048 (46.7\%) & 126,547 (86.5\%) \\
    Test & 10,522 (53.1\%) & 9,302 (46.9\%) & 19,824 (13.5\%) \\ 
    \hline
    Total & 78,021 (53.3\%) & 68,350 (46.7\%) & \textbf{146,371} \\
    \hline
  \end{tabular}
  \label{tab:split} 
\end{table}
}

%%%%%%%%%%%%%%%%%%%%%%%%%%%%%%%%%%%%%%%%%%%%%%%%%
% Feature Counts for selected dataset
%%%%%%%%%%%%%%%%%%%%%%%%%%%%%%%%%%%%%%%%%%%%%%%%%
\newcommand{\tableFeatureCount}
{
\begin{table}[htbp]
  \centering 
  \caption{Statistics of the selected datasets. Number of observations refers to the $10$-year window under study (2008-2017) for the final cohort. Raw features correspond to the number of columns in the used raw database. Input features corresponds to the features \emph{for each year} from the dataset fed into the model. The last column corresponds to the average number of non zero values among yearly feature vectors.} 
  \resizebox{\textwidth}{!}{
  \begin{tabular}{|c|c|c|c|c|}
  \hline
    Dataset & \#Obs & \#Raw Features & \#Input Features & Avg \# non-zero \\ 
    \hline
    Fixed Features (RPDB) & - & 36 & 240 & 18.76 \\
    Chronic Diseases & - &  12 & 12 & 0.54 \\
    OLIS & 512,593 & 6 & 34 & 0.65 \\
    DAD & 446,890 & 16 & 305 & 1.77 \\
    NACRS & 852,374 & 12 & 63 & 1.19 \\
    ODB & 13,857,648 & 4 & 41 & 1.08 \\
    OHIP & 37,701,635 & 4 & 153 & 18.27 \\
    ERCLAIM & 857,249 & 6 & 115 & 1.85 \\
    \hline 
    Total  & - & 96 & 963 & 44.11 \\
    \hline
  \end{tabular}}
  \label{tab:feature} 
\end{table}
}

%%%%%%%%%%%%%%%%%%%%%%%%%%%%%%%%%%%%%%%%%%%%%%%%%
% Dataset Coverage
%%%%%%%%%%%%%%%%%%%%%%%%%%%%%%%%%%%%%%%%%%%%%%%%%
\newcommand{\tableDatasetCoverage}
{
\begin{table}[!h]
  \centering 
  \caption{Dataset Descriptions. Note that a patient can have multiple observations in a dataset. Diabetes (resp. control)  coverage correspond to the percentage of the diabetes (resp. control) cohort is present in the dataset.\newline \scriptsize{(RPDB = Registered Persons Database; CHF = Congestive Heart Failure; COPD = Chronic Obstructive Pulmonary Disease; HYPER = Hypertension; OCCC = Ontario Crohn’s and Colitis Cohort; ORAD = Ontario Rheumatoid Arthritis Dataset; OMID = Ontario Myocardial Infarction Dataset; ADP = Assistive Devices Program; OCCI = Ontario Case Costing Initiative; OTR = Ontario Trauma Registry; OLIS = Ontario Laboratories Information System; LOC = Levels of Care Classification System; CFDR = Canadian Cystic Fibrosis Data Registry; CCRS = Continuing Care Reporting System; DAD = Discharge Abstract Database; NACRS = National Ambulatory Care Reporting System; ODB = Ontario Drug Benefit Claims; OHIP = Ontario Health Insurance Plan Claims Database; OMHRS = Ontario Mental Health Reporting System; ERCLAIM = OHIP's Emergency Claims Database; NRS = National Rehabilitation Reporting System)}
  } 
  \resizebox{\textwidth}{!}{
  \begin{tabular}{|c|c|c|c|c|}\hline
    Dataset & Available since & Cohort Coverage (\%) & Diabetic coverage (\%) & Control coverage (\%) \\ \hline
    \multicolumn{5}{|c|}{Base}\\ \hline
    RPDB & 1991 & 100 & 100 & 100\\
    \hline
    \multicolumn{5}{|c|}{Chronic Diseases}\\ \hline
Asthma & 1993 & 12.64 & 15.18 & 9.73 \\
CHF    & 1994 & 6.59  & 9.19  & 3.63  \\
COPD   & 1996 & 15.14 & 18.23 & 11.61 \\
Hyper  & 1991 & 48.73 & 61.96 & 33.64 \\
OCCC   & 1994 & 0.79  & 0.89  & 0.67  \\
ORAD   & 1996 & 1.60  & 1.77  & 1.40 \\ \hline
\multicolumn{5}{|c|}{Observations} \\ \hline
OMID          & 1992 & 2.40  & 3.37  & 1.30  \\
ADP - INSULIN & 2000 & 0.07  & 0.13  & 0.00  \\
OTR           & 2009 & 0.30  & 0.36  & 0.23  \\
OLIS          & 2006 & 45.92 & 51.57 & 39.47 \\
LOC           & 1997 & 0.00  & 0.00  & 0.00  \\
CFDR          & 1993 & 0.00  & 0.00  & 0.00  \\
CCRS          & 1996 & 1.47  & 1.99  & 0.87  \\
DAD           & 1989 & 52.75 & 58.21 & 46.52 \\
NACRS         & 1999 & 61.88 & 68.10 & 54.79 \\
ODB           & 1990 & 38.9 & 45.74 & 31.09 \\
OHIP          & 1990 & 87.64 & 92.95 & 81.58 \\
OMHRS         & 2005 & 1.8  & 2.32  & 1.19  \\
ERCLAIM       & 1991 & 60.47 & 66.54 & 53.55 \\
NRS           & 2000 & 0.25 & 0.11 & 0.40 \\ \hline
  \end{tabular}}
  \label{tab:dataset} 
\end{table}
}

%%%%%%%%%%%%%%%%%%%%%%%%%%%%%%%%%%%%%%%%%%%%%%%%%
% Dataset Description
%%%%%%%%%%%%%%%%%%%%%%%%%%%%%%%%%%%%%%%%%%%%%%%%%
\newcommand{\tableDatasetDescription}
{
\begin{table}[htbp]
\centering
\caption{Dataset Descriptions}
\resizebox{\textwidth}{!}{
\begin{tabular}{|c|c|c|c|c|}
\hline
Dataset & Full Name & \#Obs. & Available Time & Description \\
\hline
OLIS & Ontario Laboratories Information System & 76,009,197 & 2006-- & Lab Values\\
DAD & Discharge Abstract Database & 23,699,596 & 1989-- &  Hospitalization Records\\
NACRS & National Ambulatory Care Reporting System & 44,811,484 & 1999-- & Ambulatory Records\\
ODB & Ontario Drug Benefit Claims & 1,412,758,247 & 1990-- & Drugs Claims\\
OHIP & Ontario Health Insurance Plan Claims Database & 1,855,101,924 & 1990-- & Insurance Claims\\
ERCLAIM & OHIP's Emergency Claims Database & 37,692,110 & 1991-- & Emergency Claims\\
\hline
\end{tabular}}
\label{tab:dataset-description}
\end{table}
}

%%%%%%%%%%%%%%%%%%%%%%%%%%%%%%%%%%%%%%%%%%%%%%%%%
% Dataset Description
%%%%%%%%%%%%%%%%%%%%%%%%%%%%%%%%%%%%%%%%%%%%%%%%%
\newcommand{\tableThresholds}
{
\begin{table}[htbp]
\centering
\caption{Categorical Thresholds. The threshold is the percentage of the number of rows}
\begin{tabular}{|c|c|c|}
\hline
Dataset & Categorical Threshold (\%) & Meta-Categorical Threshold (\%) \\
\hline
ODD & 0.5 & 0.01 \\
OLIS & 0.1 & 0.05 \\
DAD & 0.5 & 0.05 \\
NACRS & 0.5 & 0.05 \\
ODB & 0.5 & 0.05 \\
OHIP & 0.5 & 0.05 \\
ERCLAIM & 0.5 & 0.05\\
\hline
\end{tabular}
\end{table}
}

%%%%%%%%%%%%%%%%%%%%%%%%%%%%%%%%%%%%%%%%%%%%%%%%%
% CNN-LSTM
%%%%%%%%%%%%%%%%%%%%%%%%%%%%%%%%%%%%%%%%%%%%%%%%%
\newcommand{\figureCNNLSTM}{
\begin{figure}[htb]
\includegraphics[width=\textwidth]{figs/cnn_lstm}
\end{figure}
}

%%%%%%%%%%%%%%%%%%%%%%%%%%%%%%%%%%%%%%%%%%%%%%%%%
% LSTM Seq2Seq
%%%%%%%%%%%%%%%%%%%%%%%%%%%%%%%%%%%%%%%%%%%%%%%%%
\newcommand{\figureSeqSeq}{
\begin{figure}[htb]
\includegraphics[width=\textwidth]{figs/lstm_seq2seq}
\end{figure}
}

%%%%%%%%%%%%%%%%%%%%%%%%%%%%%%%%%%%%%%%%%%%%%%%%%
% Highway Unbalanced
%%%%%%%%%%%%%%%%%%%%%%%%%%%%%%%%%%%%%%%%%%%%%%%%%
\newcommand{\figureHighwayUnbalanced}
{
\begin{figure}[!h]
\pgfplotstableread[row sep=\\,col sep=&]{
    interval & neg & pos \\
0--0.05&	181	&	1	\\
0.05--0.1&	82	&	1	\\
0.1--0.15&	157	&	4	\\
0.15--0.2&	196	&	5	\\
0.2--0.25&	192	&	3	\\
0.25--0.3&	201	&	8	\\
0.3--0.35&	199	&	7	\\
0.35--0.4&	202	&	9	\\
0.4--0.45&	167	&	14	\\
0.45--0.5&	135 &	7	\\
0.5--0.55&	104	&	10	\\
0.55--0.6&	101	&	19	\\
0.6--0.65&	89	&	10	\\
0.65--0.7&	63	&	11	\\
0.7--0.75&	70	&	11	\\
0.75--0.8&	65	&	9	\\
0.8--0.85&	47	&	16	\\
0.85--0.9&	47	&	26	\\
0.9--0.95&	40  &	19	\\
0.95--1&	14	&	11	\\
    }\mydata
\centering
\caption{Predicted Probability by Highway-avg ($w=10, b=1$) for diabetes and control individuals for unbalanced test set.}
\begin{tikzpicture}
   \begin{axis}[
   ybar,
   xlabel={Predicted probability},
   ylabel={Number of individuals},
   x label style = {yshift = 0cm},
   ymin=0,
   xtick=data,
   xtick pos = bottom,
   ytick pos = left,
   bar width=0.2cm,
   enlarge x limits = 0.03,
   legend pos=north west,
   x=0.65cm,
   symbolic x coords={0--0.05, 0.05--0.1, 0.1--0.15, 0.15--0.2, 0.2--0.25, 0.25--0.3, 0.3--0.35, 0.35--0.4,0.4--0.45,0.45--0.5, 0.5--0.55, 0.55--0.6,0.6--0.65, 0.65--0.7, 0.7--0.75, 0.75--0.8,  0.8--0.85, 0.85--0.9, 0.9--0.95, 0.95--1},
   x tick label style={rotate=90, anchor=east, align=right},
   ]
        \addplot table[x=interval,y=neg]{\mydata};
        \addlegendentry{diabetes};
        \addplot table[x=interval,y=pos]{\mydata};
        \addlegendentry{control};
    \end{axis}
\end{tikzpicture}
\end{figure}
}

%%%%%%%%%%%%%%%%%%%%%%%%%%%%%%%%%%%%%%%%%%%%%%%%%
% Highway Balanced
%%%%%%%%%%%%%%%%%%%%%%%%%%%%%%%%%%%%%%%%%%%%%%%%%
\newcommand{\figureHighwayBalanced}
{
\begin{figure}[!h]
\pgfplotstableread[row sep=\\,col sep=&]{
    interval & neg & pos \\
0--0.05  &709&20\\
0.05--0.1&288&42\\
0.1--0.15&582&115\\
0.15--0.2&732&206\\
0.2--0.25&804&321\\
0.25--0.3&858&378\\
0.3--0.35&811&424\\
0.35--0.4&768&490\\
0.4--0.45&673&564\\
0.45--0.5&583&566\\
0.5--0.55&421&563\\
0.55--0.6&390&641\\
0.6--0.65&334&608\\
0.65--0.7&294&660\\
0.7--0.75&274&670\\
0.75--0.8&233&736\\
0.8--0.85&185&825\\
0.85--0.9&178&940\\
0.9--0.95&127&1068\\
0.95--1  &58 &685\\
    }\mydata
\centering
\caption{Predicted Probability by Highway-avg ($w=10, b=1$) for diabetes and control individuals.}
\begin{tikzpicture}
   \begin{axis}[
   ybar,
   xlabel={Predicted probability},
   ylabel={Number of individuals},
   x label style = {yshift = 0cm},
   ymin=0,
   xtick=data,
   xtick pos = bottom,
   ytick pos = left,
   bar width=0.2cm,
   enlarge x limits = 0.03,
   legend pos=north west,
   x=0.65cm,
   symbolic x coords={0--0.05, 0.05--0.1, 0.1--0.15, 0.15--0.2, 0.2--0.25, 0.25--0.3, 0.3--0.35, 0.35--0.4,0.4--0.45,0.45--0.5, 0.5--0.55, 0.55--0.6,0.6--0.65, 0.65--0.7, 0.7--0.75, 0.75--0.8,  0.8--0.85, 0.85--0.9, 0.9--0.95, 0.95--1},
   x tick label style={rotate=90, anchor=east, align=right},
   ]
        \addplot table[x=interval,y=neg]{\mydata};
        \addlegendentry{diabetes};
        \addplot table[x=interval,y=pos]{\mydata};
        \addlegendentry{control};
    \end{axis}
\end{tikzpicture}
\end{figure}
}

%%%%%%%%%%%%%%%%%%%%%%%%%%%%%%%%%%%%%%%%%%%%%%%%%
% LR Unbalanced
%%%%%%%%%%%%%%%%%%%%%%%%%%%%%%%%%%%%%%%%%%%%%%%%%
\newcommand{\figureLRUnbalanced}
{
\begin{figure}[!h]
\pgfplotstableread[row sep=\\,col sep=&]{
    interval & neg & pos \\
0--0.05&	79	&	0	\\
0.05--0.1&	84	&	0	\\
0.1--0.15&	36	&	4	\\
0.15--0.2&	90	&	4   \\
0.2--0.25&	200	&	10	\\
0.25--0.3&	324	&	8	\\
0.3--0.35&	311	&	6	\\
0.35--0.4&	242	&	10	\\
0.4--0.45&	212	&	11	\\
0.45--0.5&	156	&	17	\\
0.5--0.55&	147	&	16	\\
0.55--0.6&	94	&	8	\\
0.6--0.65&	80	&	8	\\
0.65--0.7&	74	&	8	\\
0.7--0.75&	54	&	23	\\
0.75--0.8&	40	&	19	\\
0.8--0.85&	39	&	13	\\
0.85--0.9&	40	&	14	\\
0.9--0.95&	26	&	16	\\
0.95--1&	24	&	10	\\
    }\mydata
\centering
\caption{Predicted Probability by LR-avg ($w=5, b=1$) for diabetes and control individuals for unbalanced test set.}
\begin{tikzpicture}
   \begin{axis}[
   ybar,
   xlabel={Predicted probability},
   ylabel={Number of individuals},
   x label style = {yshift = 0cm},
   ymin=0,
   xtick=data,
   xtick pos = bottom,
   ytick pos = left,
   bar width=0.2cm,
   enlarge x limits = 0.03,
   legend pos=north west,
   x=0.65cm,
   symbolic x coords={0--0.05, 0.05--0.1, 0.1--0.15, 0.15--0.2, 0.2--0.25, 0.25--0.3, 0.3--0.35, 0.35--0.4,0.4--0.45,0.45--0.5, 0.5--0.55, 0.55--0.6,0.6--0.65, 0.65--0.7, 0.7--0.75, 0.75--0.8,  0.8--0.85, 0.85--0.9, 0.9--0.95, 0.95--1},
   x tick label style={rotate=90, anchor=east, align=right},
   ]
        \addplot table[x=interval,y=neg]{\mydata};
        \addlegendentry{diabetes};
        \addplot table[x=interval,y=pos]{\mydata};
        \addlegendentry{control};
    \end{axis}
\end{tikzpicture}
\end{figure}
}

%%%%%%%%%%%%%%%%%%%%%%%%%%%%%%%%%%%%%%%%%%%%%%%%%
% LR Balanced
%%%%%%%%%%%%%%%%%%%%%%%%%%%%%%%%%%%%%%%%%%%%%%%%%
\newcommand{\figureLRBalanced}
{
\begin{figure}[!h]
\pgfplotstableread[row sep=\\,col sep=&]{
    interval & neg & pos \\
0--0.05&	307	&	9	\\
0.05--0.1&	333	&	14	\\
0.1--0.15&	141	&	25	\\
0.15--0.2&	323	&	53	\\
0.2--0.25&	799	&	217	\\
0.25--0.3&	1262&	479	\\
0.3--0.35&	1205&	602	\\
0.35--0.4&	1041&	666	\\
0.4--0.45&	828	&	627	\\
0.45--0.5&	640	&	708	\\
0.5--0.55&	525	&	758	\\
0.55--0.6&	450 &	751	\\
0.6--0.65&	347 &	732	\\
0.65--0.7&	282 &	619	\\
0.7--0.75&	206	&	695 \\
0.75--0.8&	144	&	617	\\
0.8--0.85&	148	&	634	\\
0.85--0.9&	148 &	660	\\
0.9--0.95&	96	&	708	\\
0.95--1&	77	&	948	\\
    }\mydata
\centering
\caption{Predicted Probability by LR-avg on averaging temporal data ($w=5, b=1$) for diabetes and control individuals.}
\begin{tikzpicture}
   \begin{axis}[
   ybar,
   xlabel={Predicted probability},
   ylabel={Number of individuals},
   x label style = {yshift = 0cm},
   ymin=0,
   xtick=data,
   xtick pos = bottom,
   ytick pos = left,
   bar width=0.2cm,
   enlarge x limits = 0.03,
   legend pos=north west,
   x=0.65cm,
   symbolic x coords={0--0.05, 0.05--0.1, 0.1--0.15, 0.15--0.2, 0.2--0.25, 0.25--0.3, 0.3--0.35, 0.35--0.4,0.4--0.45,0.45--0.5, 0.5--0.55, 0.55--0.6,0.6--0.65, 0.65--0.7, 0.7--0.75, 0.75--0.8,  0.8--0.85, 0.85--0.9, 0.9--0.95, 0.95--1},
   x tick label style={rotate=90, anchor=east, align=right},
   ]
        \addplot table[x=interval,y=neg]{\mydata};
        \addlegendentry{diabetes};
        \addplot table[x=interval,y=pos]{\mydata};
        \addlegendentry{control};
    \end{axis}
\end{tikzpicture}
\end{figure}
}

%%%%%%%%%%%%%%%%%%%%%%%%%%%%%%%%%%%%%%%%%%%%%%%%%
% XGBoost Unbalanced Updated (2:56pm Mar 25)
%%%%%%%%%%%%%%%%%%%%%%%%%%%%%%%%%%%%%%%%%%%%%%%%%
\newcommand{\figureXGBUnbalanced}
{
\begin{figure}[!h]
\pgfplotstableread[row sep=\\,col sep=&]{
    interval & neg & pos \\
0--0.05&159	&0	\\
0.05--0.1&	32	&0	\\
0.1--0.15&	133	&1	\\
0.15--0.2&	184	&3	\\
0.2--0.25&	216	&1	\\
0.25--0.3&	220	&7	\\
0.3--0.35&	218	&4	\\
0.35--0.4&	170	&8	\\
0.4--0.45&	161	&12	\\
0.45--0.5&	163	&8	\\
0.5--0.55&	160	&16 \\
0.55--0.6&	123	&13	\\
0.6--0.65&	92	&15	\\
0.65--0.7&	93	&14 \\
0.7--0.75&	82	&20	\\
0.75--0.8&	63	&19	\\
0.8--0.85&	42  &20	\\
0.85--0.9&	32	&24	\\
0.9--0.95&	9	&13	\\
0.95--1&	0	&0	\\
    }\mydata
\centering
\caption{Predicted Probability by XGBoost-avg ($w=5, b=1$) for diabetes and control individuals for unbalanced test set.}
\begin{tikzpicture}
   \begin{axis}[
   ybar,
   xlabel={Predicted probability},
   ylabel={Number of individuals},
   x label style = {yshift = 0cm},
   ymin=0,
   xtick=data,
   xtick pos = bottom,
   ytick pos = left,
   bar width=0.2cm,
   enlarge x limits = 0.03,
   legend pos=north west,
   x=0.65cm,
   symbolic x coords={0--0.05, 0.05--0.1, 0.1--0.15, 0.15--0.2, 0.2--0.25, 0.25--0.3, 0.3--0.35, 0.35--0.4,0.4--0.45,0.45--0.5, 0.5--0.55, 0.55--0.6,0.6--0.65, 0.65--0.7, 0.7--0.75, 0.75--0.8,  0.8--0.85, 0.85--0.9, 0.9--0.95, 0.95--1},
   x tick label style={rotate=90, anchor=east, align=right},
   ]
        \addplot table[x=interval,y=neg]{\mydata};
        \addlegendentry{diabetes};
        \addplot table[x=interval,y=pos]{\mydata};
        \addlegendentry{control};
    \end{axis}
\end{tikzpicture}
\end{figure}
}
%%%%%%%%%%%%%%%%%%%%%%%%%%%%%%%%%%%%%%%%%%%%%%%%%
% XGBoost Balanced Updated (2.53pm mar 25)
%%%%%%%%%%%%%%%%%%%%%%%%%%%%%%%%%%%%%%%%%%%%%%%%%
\newcommand{\figureXGBBalanced}
{
\begin{figure}[!h]
\pgfplotstableread[row sep=\\,col sep=&]{
    interval & neg & pos \\
0--0.05	    &	622	&	17	\\
0.05--0.1	&	126	&	8	\\
0.1--0.15	&	488	&	56	\\
0.15--0.2	&	768	&	151	\\
0.2--0.25	&	816	&	216	\\
0.25--0.3	&	861	&	334	\\
0.3--0.35	&	828	&	429	\\
0.35--0.4	&	769	&	565	\\
0.4--0.45	&	738	&	566	\\
0.45--0.5	&	633	&	582	\\
0.5--0.55	&	527	&	585	\\
0.55--0.6	&	477	&	626	\\
0.6--0.65	&	419	&	683	\\
0.65--0.7	&	376	&	801	\\
0.7--0.75	&	281	&	919	\\
0.75--0.8	&	239	&	928	\\
0.8--0.85	&	174	&	1050	\\
0.85--0.9	&	112	&	1108	\\
0.9--0.95	&	45	&	856	\\
0.95--1	    &	3	&	42	\\
    }\mydata
\centering
\caption{Predicted Probability by XGBoost-avg ($w=5, b=1$) for diabetes and control individuals.}
\begin{tikzpicture}
   \begin{axis}[
   ybar,
   xlabel={Predicted probability},
   ylabel={Number of individuals},
   x label style = {yshift = 0cm},
   ymin=0,
   xtick=data,
   xtick pos = bottom,
   ytick pos = left,
   bar width=0.2cm,
   enlarge x limits = 0.03,
   legend pos=north west,
   x=0.65cm,
   symbolic x coords={0--0.05, 0.05--0.1, 0.1--0.15, 0.15--0.2, 0.2--0.25, 0.25--0.3, 0.3--0.35, 0.35--0.4,0.4--0.45,0.45--0.5, 0.5--0.55, 0.55--0.6,0.6--0.65, 0.65--0.7, 0.7--0.75, 0.75--0.8,  0.8--0.85, 0.85--0.9, 0.9--0.95, 0.95--1},
   x tick label style={rotate=90, anchor=east, align=right},
   ]
        \addplot table[x=interval,y=neg]{\mydata};
        \addlegendentry{diabetes};
        \addplot table[x=interval,y=pos]{\mydata};
        \addlegendentry{control};
    \end{axis}
\end{tikzpicture}
\end{figure}
}

%%%%%%%%%%%%%%%%%%%%%%%%%%%%%%%%%%%%%%%%%%%%%%%%%
% CNN-LSTM Balanced Updated 
%%%%%%%%%%%%%%%%%%%%%%%%%%%%%%%%%%%%%%%%%%%%%%%%%
\newcommand{\figureCNNBalanced}
{
\begin{figure}[!h]
\pgfplotstableread[row sep=\\,col sep=&]{
    interval & neg & pos \\
0--0.05	    &	910	&	44	\\
0.05--0.1	&	451	&	97	\\
0.1--0.15	&	648	&	172	\\
0.15--0.2	&	687	&	225	\\
0.2--0.25	&	648	&	267	\\
0.25--0.3	&	629	&	291	\\
0.3--0.35	&	551	&	293	\\
0.35--0.4	&	503	&	307	\\
0.4--0.45	&	485	&	339	\\
0.45--0.5	&	452	&	370	\\
0.5--0.55	&	414	&	429	\\
0.55--0.6	&	381	&	502	\\
0.6--0.65	&	391	&	483	\\
0.65--0.7	&	386	&	524	\\
0.7--0.75	&	359	&	614	\\
0.75--0.8	&	368	&	746	\\
0.8--0.85	&	338	&	997	\\
0.85--0.9	&	353	&	1296	\\
0.9--0.95	&	268 &	1508	\\
0.95--1	    &	80	&	1018	\\
    }\mydata
\centering
\caption{Predicted Probability by CNN-LSTM ($w=10, b=1$) for diabetes and control individuals.}
\begin{tikzpicture}
   \begin{axis}[
   ybar,
   xlabel={Predicted probability},
   ylabel={Number of individuals},
   x label style = {yshift = 0cm},
   ymin=0,
   xtick=data,
   xtick pos = bottom,
   ytick pos = left,
   bar width=0.2cm,
   enlarge x limits = 0.03,
   legend pos=north west,
   x=0.65cm,
   symbolic x coords={0--0.05, 0.05--0.1, 0.1--0.15, 0.15--0.2, 0.2--0.25, 0.25--0.3, 0.3--0.35, 0.35--0.4,0.4--0.45,0.45--0.5, 0.5--0.55, 0.55--0.6,0.6--0.65, 0.65--0.7, 0.7--0.75, 0.75--0.8,  0.8--0.85, 0.85--0.9, 0.9--0.95, 0.95--1},
   x tick label style={rotate=90, anchor=east, align=right},
   ]
        \addplot table[x=interval,y=neg]{\mydata};
        \addlegendentry{diabetes};
        \addplot table[x=interval,y=pos]{\mydata};
        \addlegendentry{control};
    \end{axis}
\end{tikzpicture}
\end{figure}
}

%%%%%%%%%%%%%%%%%%%%%%%%%%%%%%%%%%%%%%%%%%%%%%%%%
% CNN-LSTM Unbalanced Updated 
%%%%%%%%%%%%%%%%%%%%%%%%%%%%%%%%%%%%%%%%%%%%%%%%%
\newcommand{\figureCNNUnbalanced}
{
\begin{figure}[!h]
\pgfplotstableread[row sep=\\,col sep=&]{
    interval & neg & pos \\
0--0.05	    &	238	&	2	\\
0.05--0.1	&	135	&	2	\\
0.1--0.15	&	173	&	5	\\
0.15--0.2	&	157	&	9	\\
0.2--0.25	&	160	&	8	\\
0.25--0.3	&	167	&	4	\\
0.3--0.35	&	120	&	5	\\
0.35--0.4	&	131	&	8	\\
0.4--0.45	&	122	&	6	\\
0.45--0.5	&	114	&	9	\\
0.5--0.55	&	100	&	8	\\
0.55--0.6	&	96	&	8	\\
0.6--0.65	&	91	&	7	\\
0.65--0.7	&	85	&	17	\\
0.7--0.75	&	99	&	3	\\
0.75--0.8	&	107	&	12	\\
0.8--0.85	&	69	&	21	\\
0.85--0.9	&	97	&	26	\\
0.9--0.95	&	66  &	29	\\
0.95--1	    &	25	&	12	\\
    }\mydata
\centering
\caption{Predicted Probability by CNN-LSTM ($w=10, b=1$) for diabetes and control individuals for unbalanced test set.}
\begin{tikzpicture}
   \begin{axis}[
   ybar,
   xlabel={Predicted probability},
   ylabel={Number of individuals},
   x label style = {yshift = 0cm},
   ymin=0,
   xtick=data,
   xtick pos = bottom,
   ytick pos = left,
   bar width=0.2cm,
   enlarge x limits = 0.03,
   legend pos=north west,
   x=0.65cm,
   symbolic x coords={0--0.05, 0.05--0.1, 0.1--0.15, 0.15--0.2, 0.2--0.25, 0.25--0.3, 0.3--0.35, 0.35--0.4,0.4--0.45,0.45--0.5, 0.5--0.55, 0.55--0.6,0.6--0.65, 0.65--0.7, 0.7--0.75, 0.75--0.8,  0.8--0.85, 0.85--0.9, 0.9--0.95, 0.95--1},
   x tick label style={rotate=90, anchor=east, align=right},
   ]
        \addplot table[x=interval,y=neg]{\mydata};
        \addlegendentry{diabetes};
        \addplot table[x=interval,y=pos]{\mydata};
        \addlegendentry{control};
    \end{axis}
\end{tikzpicture}
\end{figure}
}

%%%%%%%%%%%%%%%%%%%%%%%%%%%%%%%%%%%%%%%%%%%%%%%%%
% Lstm Seq2Seq Balanced Updated 
%%%%%%%%%%%%%%%%%%%%%%%%%%%%%%%%%%%%%%%%%%%%%%%%%
\newcommand{\figureLSTMBalanced}
{
\begin{figure}[!h]
\pgfplotstableread[row sep=\\,col sep=&]{
    interval & neg & pos \\
0--0.05	    &	746	&	19	\\
0.05--0.1	&	277	&	43	\\
0.1--0.15	&	530	&	115	\\
0.15--0.2	&	690	&	185	\\
0.2--0.25	&	866	&	302	\\
0.25--0.3	&	767	&	359	\\
0.3--0.35	&	731 &	392	\\
0.35--0.4	&	727	&	460	\\
0.4--0.45	&	646	&	505	\\
0.45--0.5	&	543 &	534	\\
0.5--0.55	&	463	&	569	\\
0.55--0.6	&	425	&	633	\\
0.6--0.65	&	368	&	589	\\
0.65--0.7	&	342	&	642	\\
0.7--0.75	&	294	&	762	\\
0.75--0.8	&	261	&	770	\\
0.8--0.85	&	244	&	976	\\
0.85--0.9	&	210	&	977	\\
0.9--0.95	&	122 &	1002	\\
0.95--1	    &	50	&	688	\\
    }\mydata
\centering
\caption{Predicted Probability by LSTM-Seq2Seq ($w=10, b=1$) for diabetes and control individuals.}
\begin{tikzpicture}
   \begin{axis}[
   ybar,
   xlabel={Predicted probability},
   ylabel={Number of individuals},
   x label style = {yshift = 0cm},
   ymin=0,
   xtick=data,
   xtick pos = bottom,
   ytick pos = left,
   bar width=0.2cm,
   enlarge x limits = 0.03,
   legend pos=north west,
   x=0.65cm,
   symbolic x coords={0--0.05, 0.05--0.1, 0.1--0.15, 0.15--0.2, 0.2--0.25, 0.25--0.3, 0.3--0.35, 0.35--0.4,0.4--0.45,0.45--0.5, 0.5--0.55, 0.55--0.6,0.6--0.65, 0.65--0.7, 0.7--0.75, 0.75--0.8,  0.8--0.85, 0.85--0.9, 0.9--0.95, 0.95--1},
   x tick label style={rotate=90, anchor=east, align=right},
   ]
        \addplot table[x=interval,y=neg]{\mydata};
        \addlegendentry{diabetes};
        \addplot table[x=interval,y=pos]{\mydata};
        \addlegendentry{control};
    \end{axis}
\end{tikzpicture}
\end{figure}
}

%%%%%%%%%%%%%%%%%%%%%%%%%%%%%%%%%%%%%%%%%%%%%%%%%
% Lstm Seq2Seq Unbalanced Updated 
%%%%%%%%%%%%%%%%%%%%%%%%%%%%%%%%%%%%%%%%%%%%%%%%%
\newcommand{\figureLSTMUnbalanced}
{
\begin{figure}[!h]
\pgfplotstableread[row sep=\\,col sep=&]{
    interval & neg & pos \\
0--0.05	    &	200	&	1	\\
0.05--0.1	&	102	&	0	\\
0.1--0.15	&	112	&	2	\\
0.15--0.2	&	147	&	0	\\
0.2--0.25	&	154	&	4	\\
0.25--0.3	&	164	&	8	\\
0.3--0.35	&	180	&	4	\\
0.35--0.4	&	165	&	5	\\
0.4--0.45	&	160	&	9	\\
0.45--0.5	&	141	&	15	\\
0.5--0.55	&	121	&	6	\\
0.55--0.6	&	133	&	17	\\
0.6--0.65	&	99	&	12	\\
0.65--0.7	&	112 &	12	\\
0.7--0.75	&	107 &	11	\\
0.75--0.8	&	69	&	15	\\
0.8--0.85	&	76	&	19	\\
0.85--0.9	&	63	&	17	\\
0.9--0.95	&	31  &	30	\\
0.95--1	    &	16	&	7	\\
    }\mydata
\centering
\caption{Predicted Probability by LSTM-Seq2Seq ($w=10, b=1$) for diabetes and control individuals for unbalanced test set.}
\begin{tikzpicture}
   \begin{axis}[
   ybar,
   xlabel={Predicted probability},
   ylabel={Number of individuals},
   x label style = {yshift = 0cm},
   ymin=0,
   xtick=data,
   xtick pos = bottom,
   ytick pos = left,
   bar width=0.2cm,
   enlarge x limits = 0.03,
   legend pos=north west,
   x=0.65cm,
   symbolic x coords={0--0.05, 0.05--0.1, 0.1--0.15, 0.15--0.2, 0.2--0.25, 0.25--0.3, 0.3--0.35, 0.35--0.4,0.4--0.45,0.45--0.5, 0.5--0.55, 0.55--0.6,0.6--0.65, 0.65--0.7, 0.7--0.75, 0.75--0.8,  0.8--0.85, 0.85--0.9, 0.9--0.95, 0.95--1},
   x tick label style={rotate=90, anchor=east, align=right},
   ]
        \addplot table[x=interval,y=neg]{\mydata};
        \addlegendentry{diabetes};
        \addplot table[x=interval,y=pos]{\mydata};
        \addlegendentry{control};
    \end{axis}
\end{tikzpicture}
\end{figure}
}

%%%%%%%%%%%%%%%%%%%%%%%%%%%%%%%%%%%%%%%%%%%%%%%%%
% Highway Architecture
%%%%%%%%%%%%%%%%%%%%%%%%%%%%%%%%%%%%%%%%%%%%%%%%%
\newcommand{\figureHighwayArchitecture}
{
\begin{figure}[!h]
\centering 
\begin{minipage}{0.3\textwidth}
\centering
\caption{Highway block}
\begin{tikzpicture}
\draw (-0.25, -5.5) -- (-0.25, -4.5);
\draw (-1.25, -4.5) -- (0.75, -4.5);
\draw[->] (-1.25, -4.5) -- (-1.25, -2);
\draw[->] (0.75, -4.5) -- (0.75, -4);
\draw[rounded corners] (-0.35, -4) rectangle (1.85, -3.5);
\node at (0.75, -3.75) {FC + Relu};
%\drawbox{0.75, -3.75, FC + Relu, 1.1, 0.25}
\draw[->] (0.75, -3.5) -- (0.75, -3);
\draw[rounded corners] (-0.85, -3) rectangle (2.35, -2.5);
\node at (0.75, -2.75) {dropout ($p=0.5$)};
\draw[->] (0.75, -2.5) -- (0.75, -2);
\draw (-1.25, -1.9) circle (0.1);
\draw (0.75, -1.9) circle (0.1);
\filldraw[color=black] (-0.25, -1.1) circle (0.1);
\draw (-0.25, -1.1) -- (-1.05, -1.9);
\draw[->] (-0.25, -1) -- (-0.25, 0);
\draw[rounded corners] (-1.75, -5) rectangle (2.5, -0.5);
% \node at (1, -1) {\huge{H}};
\end{tikzpicture}
\end{minipage}%
~
\begin{minipage}{0.6\textwidth}
\centering
\caption{Our Highway Network. Note that highway block won't change the dimension of the input.}
\begin{tikzpicture}
\draw[rounded corners] (-2, -2) rectangle (2, -1.5);
\node at (0, -1.75) {input};
\draw[->] (0, -1.5) -- (0, -1);
\draw[rounded corners] (-2, -1) rectangle (2, -0.5);
\node at (0, -0.75) {FC (dim=2000)};
\draw[->] (0, -0.5) -- (0, 0);
\draw[rounded corners] (-2, 0) rectangle (2, 0.5);
\node at (0, 0.25) {Highway block};
\draw[->] (0, 0.5) -- (0, 1);
\draw[rounded corners] (-2, 1) rectangle (2, 1.5);
\node at (0, 1.25) {FC (dim=1000)};
\draw[->] (0, 1.5) -- (0, 2);
\draw[rounded corners] (-2, 2) rectangle (2, 2.5);
\node at (0, 2.25) {Highway block};
\draw[->, rounded corners] (0, 2.5) -- (0, 3) -- (2.5, 3) -- (2.5, -2.5) -- (5, -2.5) -- (5, -2);
\draw[rounded corners] (3, -2) rectangle (7, -1.5);
\node at (5, -1.75) {FC (dim=500)};
\draw[->] (5, -1.5) -- (5, -1);
\draw[rounded corners] (3, -1) rectangle (7, -0.5);
\node at (5, -0.75) {Highway block};
\draw[->] (5, -0.5) -- (5, 0);
\draw[rounded corners] (3, 0) rectangle (7, 0.5);
\node at (5, 0.25) {FC (dim=1)};
\draw[->] (5, 0.5) -- (5, 1);
\draw[rounded corners] (3, 1) rectangle (7, 1.5);
\node at (5, 1.25) {sigmoid};
\draw[->] (5, 1.5) -- (5, 2);
\draw[rounded corners] (3, 2) rectangle (7, 2.5);
\node at (5, 2.25) {output};
\end{tikzpicture}
\end{minipage}
\end{figure}
}

\newcommand{\figureXGBBufferY}
{
\begin{figure*}[htbp]
\centering
\caption{Effect of window size $w$ and buffer size $b$. }
\begin{tikzpicture}
\begin{groupplot}[
        group style={
            group size=2 by 1,
            % only set the tick and axis labels at the plots on the left
            % and on the bottom
            x descriptions at=edge bottom,
            %y descriptions at=edge left,
        }
    ]
\nextgroupplot[
  width = 0.5\textwidth,
  xlabel={window size $w$ ($b=1$)},
    legend style={font=\small, at={(0,0)}, anchor=center, legend columns=4},
    legend to name=wb,
    xtick={1,2,3,4,5,6,7,8,9,10},
    ytick={73,74,75,76,77,78,79,80,81},
    xmajorgrids,
    ymajorgrids,
  ylabel=AUC]
\addlegendentry{LR-avg}
\addplot[color=red, mark=*, thick] coordinates {
(1, 74.2)
(3, 77)
(5, 77.6)
(10, 77.7)
};
\addlegendentry{XGBoost-avg}
\addplot[color=blue, mark=*, thick] coordinates {
(1, 76.4)
(3, 79.3)
(5, 79.9)
(10, 79.7)
};
\addlegendentry{Highway-avg}
\addplot[color=black, mark=*, thick] coordinates {
(1, 74.4)
(3, 77.5)
(5, 78.2)
(10, 78.6)
};
\addlegendentry{CNN-LSTM}
\addplot[color=orange, mark=*, thick] coordinates {
(3, 76.6)
(5, 77.5)
(10, 78)
};
\addlegendentry{LR-concat}
\addplot[color=red, mark=*, dotted, thick] coordinates {
(1, 74.2)
(3, 77.1)
(5, 77.8)
(10, 78)
};
\addlegendentry{XGBoost-concat}
\addplot[color=blue, mark=*, dotted, thick] coordinates {
(1, 76.4)
(3, 79.6)
(5, 80.3)
(10, 79.7)
};
\addlegendentry{Highway-concat}
\addplot[color=black, mark=*, dotted, thick] coordinates {
(1, 74.4)
(3, 76.9)
(5, 77.2)
(10, 76.7)
};
\addlegendentry{LSTM-Seq2Seq}
\addplot[color=orange, mark=*, dotted, thick] coordinates {
(3, 76.3)
(5, 77.4)
(10, 78.4)
};
\nextgroupplot[
  width=0.5\textwidth,
      xmajorgrids,
    ymajorgrids,
  xlabel={buffer size $b$ ($w=5$)},
  xtick={1,2,3,4,5,6,7,8,9,10},
  ytick={68,69,70,71,72,73,74,75,76,77,78,79,80}
  ]
%   \addplot[color=red, mark=*, thick] coordinates {
% %(0, 78.8)
% (1, 76.5)
% (2, 75.1)
% (3, 73.7)
% (4, 72.4)
% (5, 71.1)
% (6, 69.9)
% (7, 68.6)
% (8, 67.9)
% (9, 67.3)
% (10, 67)
% };
\addplot[color=blue, mark=*, thick] coordinates {
%(0, 82.2)
(1, 79.9)
(2, 77.9)
(3, 75.9)
(4, 74.4)
(5, 72.5)
(6, 71.1)
(7, 69.4)
(8, 68.9)
(9, 68.4)
(10, 68)
};
% \addplot[color=black, mark=*, thick] coordinates {
% %(0, 80.7)
% (1, 76.9)
% (2, 75.9)
% (3, 73.7)
% (4, 72)
% (5, 70.9)
% (6, 69.2)
% (7, 69.1)
% (8, 68)
% (9, 67.8)
% (10, 67.5)
% };
\end{groupplot}
\node[above=0em] at(current bounding box.north) {\pgfplotslegendfromname{wb}};
\end{tikzpicture}
\label{fig:result-buffer}
\end{figure*}
}

%%%%%%%%%%%%%%%%%%%%%%%%%%%%%%%%%%%%%%%%%%%%%%%%%
% Study Design
%%%%%%%%%%%%%%%%%%%%%%%%%%%%%%%%%%%%%%%%%%%%%%%%%
\newcommand{\figureStudyDesign}
{
\begin{figure}[htbp]
\centering
\caption{Study design}
\begin{tikzpicture}
\filldraw[color=lightgray] (1.5, -0.5) rectangle (4, 0);
\filldraw[color=yellow] (-6, -0.5) rectangle (1.5, 0);

\draw[-,thick] (-7,-0.5)--(5,-0.5);
\draw[-,thick] (-7,0)--(5,0);
\draw[->, thick] (5.5, -0.25) -- (6.75, -0.25);

\draw[-,thick] (4, -0.5) -- (4, 0);
\draw[-,thick] (1.5, -0.5) -- (1.5, 0);
\draw[-,thick] (-6, -0.5) -- (-6, 0);
\node[above right] at (5.5,-0.25) {time};
\node[above] at (1.5, 0) {current date};
\node[above] at (4, 0) {prediction date};
\node[above] at (-6, 0) {observation start date};
\node at (2.75, -0.25) {buffer ($b$)};
\node at (-2.25, -0.25) {observation window ($w$)};

\end{tikzpicture}
\label{fig:studydesign}
\end{figure}
}

%%%%%%%%%%%%%%%%%%%%%%%%%%%%%%%%%%%%%%%%%%%%%%%%%
% Observation Window
%%%%%%%%%%%%%%%%%%%%%%%%%%%%%%%%%%%%%%%%%%%%%%%%%
\newcommand{\figureObservationWindow}
{
\begin{figure}[htbp]
\centering 
\caption{Feature vector for an individual}
\begin{tikzpicture}
\filldraw[color=yellow] (-6, -0.5) rectangle (4, 0);

\draw[-,thick] (-7,-0.5)--(5,-0.5);
\draw[-,thick] (-7,0)--(5,0);
\draw[->, thick] (5.5, -0.25) -- (6.75, -0.25);

\draw[-,ultra thick] (4, -0.5) -- (4, 0);
\foreach \i in {2,0,-2,-4}
    \draw[-] (\i, -0.5) -- (\i, 0);
\draw[-,ultra thick] (-6, -0.5) -- (-6, 0);
\node[above right] at (5.5,-0.25) {time};

\node at (-1, 0.75) {observation window};
\foreach \i in {0,...,5}
    \node[above] at (\i*2 - 6, 0) {year \i};
\foreach \i in {0,...,4} 
{
    \draw[-] (\i * 2 - 5.95, -0.75) -- (\i * 2 - 5.95, -1.25) -- (\i * 2 - 4.05, -1.25) -- (\i * 2 - 4.05, -0.75);
    \draw[->] (\i * 2 - 5, -1.25) -- (\i * 2 - 5, -1.5);
    \node[above] at (\i * 2 - 5, -1.25) {agg};
}
\foreach \i in {-5, -3, -1, 1, 3}
{
    \filldraw[color=lightgray] (\i - 0.8, -2.25) rectangle (\i + 0.8, -1.75);
    %\filldraw[color=lime] (\i - 0.8, -3.25) rectangle (\i + 0.8, -2.75);
    % \node at (\i, -2.5) {+};
    \node at (\i, -2) {temporal};
    %\node at (\i, -3) {chronic};
}
\node at (-1, -2.5) {+};
\filldraw[color=lime] (-3, -3.25) rectangle (1, -2.75);
\node at (-1, -3) {fixed features};
\draw[->, line width=5pt] (-1, -3.5) -- (-1, -4.25);

\filldraw[color=lime] (-7, -5) rectangle (-5, -4.5);
\foreach \i in {-5, -3, -1, 1, 3} 
{
    \filldraw[color=lightgray] (\i, -5) rectangle (\i + 2, -4.5);
    \draw (\i, -5) -- (\i, -4.5);
    %\filldraw[color=lime] (\i + 1.25, -6) rectangle (\i + 2, -5.5);
}
\node at (-1,-5.25) {or};
\filldraw[color=lime] (-3, -6) rectangle (-1, -5.5);
\filldraw[color=lightgray] (-1, -6) rectangle (1, -5.5);
\draw (-1, -6) -- (-1, -5.5);
%\filldraw[color=lime] (0.25, -7) rectangle (1, -6.5);

\node at (6, -4.75) {concat};
\node at (6, -5.75) {average};
\end{tikzpicture}
\label{fig:obswindow}
\end{figure}
}

\maketitle

\begin{abstract}
 % Summary of the article.  Be sure to highlight the technical significance and clinical relevance.
Leveraging health administrative data (HAD) datasets for predicting the risk of chronic diseases including diabetes has gained a lot of attention in the machine learning community recently. In this paper, we use the largest health records datasets of patients in Ontario, Canada. Provided by the Institute of Clinical Evaluative Sciences (ICES), this database is age, gender and ethnicity-diverse. The datasets include demographics, lab measurements, drug benefits, healthcare system interactions, ambulatory and hospitalizations records. We perform one of the first large-scale machine learning studies with this data to study the task of predicting diabetes in a range of $1-10$ years ahead, which requires no additional screening of individuals.
 
 In the best setup, we reach a test AUC of $80.3$ with a single-model trained on an observation window of $5$ years with a one-year buffer using all datasets. A subset of top $15$ features alone (out of a total of $963$) could provide a test AUC of $79.1$. In this paper, we provide extensive machine learning model performance and feature contribution analysis, which enables us to narrow down to the most important features useful for diabetes forecasting. Examples include chronic conditions such as asthma and hypertension, lab results, diagnostic codes in insurance claims, age and geographical information.
 
\end{abstract}

%%%%%%%%%%%%%%%%%%%%%%%%%%%%%%%%%%%%%%%%%%%%%%%%%%%%%%%%%%%%%%%%%%%%%%%%%%
\section{Introduction}
%\red{This is where we talk about diabetes. Tells us a bit about the problem.  Recent advances in machine learning \citep{cite1} have resulted in great things happening in healthcare. In particular, \citet{cite2} describes a spiffy technique to save even more lives.  In this work, we...}

%%%%%%%%%%%%%%% chronic diseases
Preventable chronic conditions such as heart diseases, stroke, cancer and type 2 diabetes mellitus (T2DM) are the main causes of morbidity and mortality as of 2019 \citep{ding2018effectiveness, wu2018type, bhardwaj2018impact}. In 2014, in the United States, $7$ out of $10$ top causes of mortality were chronic diseases. Chronic diseases formed $86\%$ of the US health care expenditures in $2010$ and costed $\$345$ billion for diabetes and pre-diabetes alone in 2012 \citep{bhardwaj2018impact}.

%%%%%%%%%%%%%%%%%%   Diabetes itself!
As a long-lasting chronic disease, diabetes occurs due to pancreas failure in producing insulin, which controls the blood sugar level ; or when the body is inefficient in using the existing insulin \citep{eljerjawi}.  Late diagnosis of diabetes could result in increased macro-vascular and capillaries difficulties risk, kidney failure, etc that might increase the healthcare cost \citep{rosella2016impact} and/or even threaten the life of the patient \citep{choi2019machine}. Hence, early prediction of diabetes is of utmost value.

The availability of clinical big data opens up a lot of opportunities to revolutionize healthcare with the use of machine learning techniques \citep{osmani2018processing}. Machine learning experts can help clinicians understand how, when and what data is desired to solve their problems \citep{joyce2018}. It has been demonstrated in several literature works that machine learning techniques applied to large health administrative datasets can benefit the areas of risk prediction, hospital readmission reduction, treatment guidance, cost reduction, etc. \citep{bhardwaj2018impact, mikalsen2019advancing, thesmar2019combining, garske2018using, miotto2016deep}. Moreover, they have been proved useful in management of chronic diseases such as diabetes mellitus and could help mitigate the disease burden \citep{bhardwaj2018impact}.

Machine learning techniques can be used in two different setups for chronic diseases prediction. In the first setup \citep{alhassan2018type, zou2018predicting}, individuals are classified as \emph{currently} being with or without diabetes. In the second setup, the task is forward prediction: predicting whether patients will get diabetes or not \emph{in the future} \citep{choi2019machine, sontag}. In this work, we tackle this latter task. 

A number of demographic (fixed) features are useful in predicting diabetes. For example, age and gender have been utilized in a lot of literature works \citep{eljerjawi, sontag}. In this work, we use (among others) country of origin \citep{rosella2012role} and geographical information such as latitude and longitude.

%\red{Mat: remove this paragraph to save space} More importantly, diabetes management has been revolutionized over recent years by the introduction of electronic health records including sensor measurements, laboratory results, pharmacy records and insurance claims \citep{sontag}, which facilitates the use of machine learning methods for prediction, diagnosis, phenotyping, etc. Measurements include continuous glucose monitoring, insulin pump data, heart rate and movement available through wristbands or watches \citep{rigla2018artificial}.

%% Insurance claims
%%%%%%%%%%%%%%%%%%%%%%%%%%%%%%%%%%%%%%%
As demonstrated in a number of literature works \citep{razavian2015population, nagata2018prediction}, health insurance claims with diagnostic codes in an International Classification of Diseases (ICD) format could contain a significant signal for anticipating diabetes. Despite the sparsity associated with such data, we found these claims to strongly help diabetes prediction, especially when linked with other immune diseases or hypertension. 

%% Lab results
%%%%%%%%%%%%%%%%%%%%%%%%%%%%%%%%%%%%%%%
Lab test data obtained from health checkups or tests conducted at hospitals also play a significant role in our diabetes prediction \citep{nagata2018prediction}.  For instance, HbA1c (or A1c for short) is proven to be useful in predicting type 2 in different common scenarios and in diagnosing individuals with elevated T2D risk in both the short and long term \citep{leong2018prediction}. We will showcase its contribution to the task of diabetes prediction further.

The single-payer healthcare system in Canada, similarly to UK and a few other countries provides a unified collection of HAD records for all residents and that makes the Institute for Clinical Evaluative Sciences (ICES) collection of datasets extremely valuable for related healthcare studies. Moreover, ICES is quite diverse in terms of gender (53-47 male-female ratio in our cohort), age (year of birth from the 1890s to 2010s) and country of birth (more than 100).

%%%%%%%%%%%% methods
In this work, we perform diabetes mellitus (type I and II combined) onset prediction using raw features from a total of 20 datasets obtained from ICES recorded in Ontario, Canada for 146,371 individuals. Among them, 78,021 individuals have been diagnosed with diabetes between 2008 and 2018. Each dataset corresponds to a different source of data, including: lab results (OLIS), insurance claims (OHIP), drug benefits (ODB), hospitalizations (DAD), ambulatory usage (NACRS) and emergency claims (ERCLAIM).  

We use these features as inputs to five different machine learning models (i.e. XGBoost \citep{xgboost}, Highway networks \citep{highway}, regularized logistic regression (LR) \citep{sontag}, CNN-LSTM \citep{tonekaboni2018prediction} and LSTM sequence to sequence (LSTM-Seq2Seq) \citep{sutskever2014sequence}) to predict diabetes onset using a variable length window of temporal data. %We consider a variable size buffer to ensure zero data leakage.

%%%%%%%%%%%%%%%%%%%%%%%%%%%%%%%%%%%%%%%
\paragraph{Technical Significance}
To the best of our knowledge, we propose the first comprehensive large-scale machine learning study on relevant datasets from ICES collection in Ontario to solve the diabetes prediction problem. In fact, we provide: a data pipeline that efficiently aggregates sparse temporal data from multiple tables with different sizes into a unique sparse representation vector, a fair comparison of five models of different natures, multiple metrics for performance evaluation, and finally an exhaustive feature selection/contribution analysis, that investigates features' contribution to the prediction of diabetes. The latter is often of lesser interest in performance-focused studies; versus in this paper, we aim to interpret all the top features in depth.

%%%%%%%%%%%%%%%%%%%%%%%%%%%%%%%%%%%%%%%
\paragraph{Clinical Relevance}

Our model predictions can be used as a risk score for evaluating future onset of diabetes. We believe that our prediction model could be deployed to better understand the distribution of diabetes (and its main contributing features) in the population. Our method does not require additional screening cost as it simply looks at all the historical health data available.

By adopting an extensive set of input features and investigating their contribution, we attempt to reverse engineer feature selection and give a machine learning model point of view on which features should be used in priority. This feature contribution-based method can be used in recommending policies in health planning and strategies to prevent the onset of diabetes among those at high risks.  Our feature contribution analysis suggests that other chronic diseases such as asthma are linked with future diabetes. It also demonstrates the importance of lab testing results in diabetes prediction.
Finally, by varying the buffer size (see study design), we show how far in advance we can forecast diabetes with a reasonable accuracy.

%%%%%%%%%%%%%%%%%%%%%%%%%%%%%%%%%%%%%%%
\subsection{Related work}

Generally, we found two categories of works based on the machine learning problems they define for prediction of diabetes: current time (diagnosis, classification) \citep{zou2018predicting, wu2018type, alhassan2018type} and forward prediction~\citep{choi2019machine, nagata2018prediction, sontag, krishnan2013early}. As we are interested in predicting the diabetes incidence in advance (latter problem), we review the relevant forward prediction work here.

Considering the diabetes-related features used in literature works, we observed two main approaches. In the first approach, a few important features are selected or engineered manually/systematically and are used for diabetes prediction or diagnosis \citep{eljerjawi, alhassan2018type}. In the second category, all available features are inputted to a machine (or deep) learning model \citep{sontag, krishnan2013early} and it is the model's responsibility to realize the important features \citep{sontag}. In this paper, we adopt the second approach.

% %%% Sontag's paper
One of the most relevant literature works is presented in \cite{sontag}. The authors propose a data-driven model for predicting type 2 diabetes in advance (as well as current time diagnosis). They consider claims, pharmacy records, healthcare utilization and laboratory results of $4.1$ million people recorded between 2005 and 2009. 
The novel risk factors emerging from their model is studied for different age groups at different stages before the onset.
They achieve an AUC (area under curve) measure of $78$ with about $900$ features for prediction of diabetes one year in advance (i.e. with a buffer of one year). 
They adopted logistic regression as the machine learning model of choice.
Similarly, in an earlier work \citep{krishnan2013early} studied the same problem with the same method. For $22$ selected features and a complete list of $1054$ features, AUCs of $75.2$ and $75.6$ were achieved, respectively, also for one year in advance prediction.

%%%% Korean paper:      Machine Learning for the Prediction of New-Onset Diabetes Mellitus during 5-Year Follow-up in Non-Diabetic Patients with Cardiovascular paper
A recent study on machine learning-based diabetes prediction is presented in \cite{choi2019machine}. A group of $8454$ individuals (with no diabetes history) were studied over a period of $5$-year followup. $28$ variables were extracted from individual's electronic medical records to train a number of models for predicting the occurrence of diabetes within the followup period (a different problem setup compared to ours). Authors confirmed that the regularized logistic regression performed the best among their models and results in an AUC of $78.0$. Finally, they performed feature selection based on information gain attribute evaluation. In fact, their work is one of the few ones that studies features contributions. Similarly to our findings (using our top model, XGBoost), HbA1c shows up as the most important feature among theirs. Both their study and ours also give strong importance to glucose values, while we found age to be of a lesser importance than them.

%%%%%%%%  Japenese paper: Prediction Models for Risk of Type-2 Diabetes Using Health Claims

The study in \cite{nagata2018prediction} confirms the usefulness of lab test data and health claim text data including ICD10 codes and pharmacy information. They used three models for diabetes prediction : L1-regularized logistic regression, XGBoost and a family of LSTM-based models. Among them, XGBoost provides the best performance with an AUC of $87.0$ when using ICD10 and medical text data for predicting diabetes, one year ahead. Note that we did not use any medical text data in this work due to the very small coverage on our population with diabetes. 

\section{Data Cohort}
In this section, we describe how our cohort was selected among numerous datasets within ICES, where some might not be relevant to diabetes prediction. 

%%%%%%%%%%%%%%%%%%%%%%%%%%%%%%%%%%%%%%%
\subsection{Cohort Selection} 

\subsubsection{Diabetes Cohort}
The Ontario Diabetes Dataset (ODD) is used to identify individuals with diabetes. The dataset contains all individuals in the Registered Persons Database (RPDB), who have been flagged as ones with diabetes based on the algorithm originally introduced by \cite{hux}. This algorithm flags Physician Service Claims (PSCs) and fee codes and hospitalizations associated with diabetes based on Hospital Discharge Abstracts (HDAs). \footnote{This part was not present in \cite{hux}. In fact, we used the 2016 version of ODD, where the algorithm is updated to what we described here. The algorithm applies to all individuals in RPDB (even before 2016)}. ODD would also include the \emph{diagnosis date} of individuals, which is determined based on the described algorithm. Details are listed in the supplementary materials. We randomly selected $10\%$ of people appearing at least once in the ODD. %This results in $213,970$ patients. %The PSCs can be obtained from the Ontario Health Insurance Plan (OHIP) database and the HDAs through the Discharge Abstract Database (DAD, formerly CIHI - Canadian Institute for Health Information). The diagnostic code for diabetes is given by the ICD9 code 250.x or ICD10 code E10-E14. An individual is flagged with diabetes if one of the following conditions is met:
%\begin{itemize}
%    \item Two OHIP (physician) claims with diabetic diagnostic code within the past 2-year period.
%    \item One hospitalization (in DAD) with diabetic diagnostic code.
%    \item A single OHIP claim with a diabetic fee code for diabetes management, insulin therapy support, diabetic management assessment. Such code is among (Q040, K029, K030, K045 and K046). \footnote{This part was not present in \cite{hux}. In fact, we used the 2016 version of ODD, where the algorithm is updated to what we described here. The algorithm applies to all individuals in RPDB (even before 2016)}
%\end{itemize}
%When compared with primary care chart data, this algorithm reaches a sensitivity of $86\%$, a specificity of $97\%$ and an $80\%$ positive predictive value in the original paper (\cite{hux}). 

% \paragraph{Diagnosis Date}
% If an individual is flagged as diabetic using the above algorithm, the \emph{diagnosis date} is set to be the earliest date of any such events, including OHIP claims, within the past 2-year window. For example, an individual with OHIP diabetic claims in 2000, 2003 and 2004 will have their diagnosis date set to 2003. An individual with an OHIP diabetic claim in 2000 and a diabetic HDA in 2001 will have their diagnosis date set to 2000.
% We randomly select $10\%$ of people appearing at least once in the ODD. This results in $213,970$ patients. 

%[\red{LAURA'S FEEDBACK NEEDED}]

\subsubsection{Control Cohort}
Since our task is a binary prediction, we need to build a negative class of individuals without diabetes. The control cohort are individuals also selected from the RPDB. It has a similar number of individuals, who are alive at the time of study and not flagged in ODD. Besides, the control cohort has the same age and gender distribution as the diabetes cohort.

%As the diabetes cohort is diverse in terms of age and gender, we make sure that the control samples we selected are also diverse in age and gender. According to the distribution of the diagnosis date for the diabetes cohort, a ``diagnosis date'' is assigned to each control individual as well, which we need for our study design. 
%We match each diabetic patient with a ``twin'': given an ODD individual, we select a person in RPDB such that (1) they are not in ODD; (2) they are alive at the diagnosis date of the corresponding ODD individual; (3) they have the same gender as the corresponding ODD individual; and (4) their dates of birth are within 1 year of that of the corresponding ODD individual. The ``diagnosis date'' of the control individual is set to be the same as their ODD ``twin''.

\subsubsection{Exclusion Criteria}
The ODD algorithm imposes certain properties for an individual (see supplementary materials) in how the diagnosis date is determined. It is not surprising that not all individuals in the ODD would satisfy these properties, as healthcare labels are not perfect. Luckily, this contributes to a small portion (4.1\%) of our diabetes cohort and thus we excluded these individuals. We also removed the individuals in the \emph{control} cohort that violate these properties. 

We also scoped down to individuals diagnosed between 2008 to 2017 (both included). As our task is to forecast diabetes onset, it would be more relevant to focus on more recent data which are cleaner and contain lab values (OLIS, see \tableref{tab:dataset-description}). %to match with the period during the which we have lab results from OLIS (See  \tableref{tab:dataset-description}). 
Lastly, we dropped the individuals who are younger than 20, mostly corresponding to type I diabetes \citep{atkinson2014type, harjutsalo2008time}. Note that our final, adult-only cohort still contains type I cases, and we predict future incidence of type I and type II combined, The resulting cohort consists of a total of 146,371 individuals, in which 78,021(53.3\%) individuals have diabetes and 68,350(46.7\%) do not.

%Based on the ODD algorithm, there should be no diabetic hospitalization code or OHIP fee code before the diagnosis date. There should also be no OHIP diabetic claim 2 years prior to diagnosis date. We found that in ODD, some individuals ($3.5\%$ of the total cohort) do violate the above rules. In these cases, we drop those individuals to make sure our data is clean enough for prediction. A substantial amount of these patients correspond to control group patients having an OHIP diabetic fee code or claim within two years of their diagnosis date. Another portion of these individuals are gestational diabetes cases. 

%Hence, we scoped down to diabetic individuals diagnosed between January 1st, 2008 and December 31st, 2017. We finally dropped individuals younger than $20$. The resulting cohort consists of $80153$ individuals flagged with diabetes and $71571$ individuals associated individuals in the control cohort.

%%%%%%%%%%%%%%%%%%%%%%%%%%%%%%%%%%%%%%%
\subsection{Dataset Selection} 

After constructing the cohort, we first extracted data about each individual from several sources (eight datasets). These data have no timestamps - we call them \emph{fixed features}. Then we selected datasets that are both relevant and useful to our study. 

The derived \emph{chronic diseases} (Asthma, CHF, COPD, HYPER, OCCC, ORAD) consist of the incidence and prevalence of the chronic disease in each year for each individual. 
We kept all the six validated chronic diseases as they are all clinically related to diabetes.

The \emph{observations} datasets contain more detailed information associated with timestamps, and are typically very sparse. Based on the coverage of the diabetes and the control cohort appearing in the dataset, we selected six observations datasets (OLIS\footnote{OLIS is a relatively new dataset and is being organized and cleaned continuously. Right now, we just have access to part of OLIS data, most of which are diabetes-related.}, DAD, NACRS, ODB, OHIP and ERCLAIM) in our final data processing setup. They cover lab values, hospitalizations, ambulance services, and drugs, health and emergency claims. These datasets and their descriptions are outlined in \tableref{tab:dataset-description} and the coverage of all the datasets are outlined in the supplementary materials.

\tableDatasetDescription

%\tableDatasetCoverage

%%%%%%%%%%%%%%%%%%%%%%%%%%%%%%%%%%%%%%%
\subsection{Feature Selection} 
%In each dataset, we selected features based on several criteria. 
%\subsubsection{Manual Selection} 
We manually selected relevant features from each dataset, as required by the data owner. We dropped features that are 1) missing from the data most of the time; 2) overlapping with other features; or 3) without a timestamp in the observations dataset 4) occurring \emph{after} the prediction date. We then performed feature engineering based on each feature's type. Details are in the supplementary materials.
Our model input is built based on a $w$-year observation window (see \sectionref{section:study-design}). As multiple observations can occur per year, we aggregated each observation feature within each year. By default, we averaged observations within each year. We found it useful to also include the maximum values of observations within each year.
We also included in all cases the number of observations per year per dataset for each individual. %Then, we computed one feature vector per year for each patient, maintaining constant feature dimension across time and patients. 

Chronic diseases are recorded yearly hence they do not require aggregation. The date of each chronic disease in a specific year is set to be January 1st of that year. We concatenated the aggregated observations and the chronic diseases as \emph{temporal vectors}. Thus, for each individual, the feature vector contains a fixed vector portion, and $w$ temporal vectors corresponding to each year. We then experimented with concatenating these $w$ yearly vectors, averaging them, or inputting them in a sequential (temporal) fashion for our LSTMs. See  \figureref{fig:obswindow} for illustration. The number of raw features and input features for each dataset is described in \tableref{tab:feature}.

\figureObservationWindow

%%%%%%%%%%%%%%%%%%%%%%%%%%%%%%%%%%%%%%%
%\subsection{Statistics}
%Table \ref{tab:feature} contains selected statistics of the features and datasets. % Note that the fixed features contain both the set of features which are for sure fixed, like birth year; and the set of features that could be changed, but in a low frequency, like location, rurality or latitude and longitude. In our study, these potentially time-dependent fixed attributes were captured at the diagnosis date. 

\tableFeatureCount

%%%%%%%%%%%%%%%%%%%%%%%%%%%%%%%%%%%%%%%
%\subsection{Implementation}
%We built our dataset extraction pipeline using Java to load, parse and aggregate all the datasets (including the ones that are not used). To ensure scalability, we processed the patients in chunks and used sparse representation for all data. \red{SHALL WE MENTION THIS?!?! We hope that our data pipeline would be of help to all individuals who wish to conduct machine learning studies with these datasets}. The output of this pipeline is in JSON format containing the patient key, label, feature vector in sparse representation and feature names. \red{HMMM It can also be nicely loaded in Python for machine learning model model training.} Finally, the models were implemented in Python using Pytorch.

%%%%%%%%%%%%%%%%%%%%%%%%%%%%%%%%%%%%%%%%%%%%%%%%%%%%%%%%%%%%%%%%%%%%%%%%%%
\section{Methods}

%%%%%%%%%%%%%%%%%%%%%%%%%%%%%%%%%%%%%%%
\subsection{Study design}\label{section:study-design}
We designed our study to predict the possibility for an individual to have diabetes in the future. We built the feature vectors from data described above by considering $w$ years before the current date. We then predicted whether the individuals were to have an onset of diabetes in $b$ years in the future. To formulate the diabetes prediction problem for the training data, we set the ``current date'' to be $b$ years before the \emph{prediction date}. For the diabetes cohort, the prediction date is the diagnosis date. For the control cohort, the prediction date is sampled to match the distribution of the diagnosis dates of the diabetes cohort (from 2008 to 2017). The design of the prediction task at an individual level is illustrated in \figureref{fig:studydesign}. In our study, $b = 1,\ldots, 10$ and $w=1,3,5,10$. If the individual does not have enough history in the observation window, we pad with zeros.
% we:
% \begin{itemize}
%     \item Consider the diagnosis date $t$ of a \red{diabetic} individual. The diagnosis date of a non-diabetic individual is outlined above in the cohort selection section.
%     \item Determine a buffer of $b$ years and an observation window of $w$ years. (Here $w$ can be variable length based on different individual.)
%     \item Our observation window will be $[t-b-w, t-b]$. Here $t-b$ would be the ``current date'' described above.
% \end{itemize}
% In our study $b = 0, \ldots, 10$ and $w = 1,3,5, 10$. If the individual does not have enough history in the observation window, we pad them with zeros. The design of the prediction task at the patient level is illustrated in Figure \ref{fig:studydesign}.

\figureStudyDesign

%%%%%%%%%%%%%%%%%%%%%%%%%%%%%%%%%%%%%%%
\subsection{Split}
As our main focus is on the clinical application, we used \emph{out-of-time} test split. As such, all the individuals with prediction date in 2016 or 2017 were set as test data, while the ones between 2008 and 2015 (all inclusive) formed the training data. The goal of deploying our model in production for future diabetes risk screening naturally led to this setup. The train-test distribution is outlined in \tableref{tab:split}.

\tableTrainTestSplit

%%%%%%%%%%%%%%%%%%%%%%%%%%%%%%%%%%%%%%%
\subsection{Models}

As depicted in \figureref{fig:obswindow}, we have two kinds of final input vectors, \emph{avg} and \emph{concat}. While \emph{concat} carries temporal information that may be helpful, it also increases the size of the feature vector $w$-fold ($w$ is the observation window size). We experimented both types of input vectors for our study below, except for LSTMs as they are designed to capture the temporal aspect. Details of these models are described in the supplementary materials.
%In our models below, except for \red{LSTM-Seq2Seq}, which captures the temporal aspects, we experimented with both types of input vectors.

\paragraph{Logistic Regression} We use an L1-regularized logistic regression (LR) model as our baseline. 

\paragraph{XGBoost}
We use XGBoost, introduced by \cite{xgboost}, as our main model due to its popularity, high performance and ability to handle missing values well.

\paragraph{Highway Network} We also use a relatively simple deep neural network without temporal aspect called the Highway network \citep{highway}. It combines the linear and non-linear outputs providing greater modeling flexibility for the network compared to the vanilla multi-layer perceptron (MLP). We found that this network is better than vanilla MLPs in our application. This neural network has 14,944,001 parameters in the case of \emph{avg} input vectors. 

\paragraph{CNN-LSTM}This model is inspired by the architecture in \cite{tonekaboni2018prediction} with a number of changes as follow: We removed the max pooling layers as we found them damaging the performance. No feature dimensionality reduction was performed in our setup. We added batch normalization \citep{ioffe2015batch} after each CNN layer, which increased the performance and convergence speed simultaneously. Dropout in the reference architecture was kept in the same place, i.e. after CNN layers. Two layers of LSTM were used. Finally, we applied a three layer perceptron followed by sigmoid to perform binary prediction. This model has a total of 22,744,136 parameters. 

\paragraph{LSTM-Seq2Seq} 
We borrowed the LSTM-based sequence to sequence architecture from \cite{sutskever2014sequence}. We used it to reconstruct the concatenation of all (fixed + temporal) features over time and applied a three layer perceptron on the encoder last hidden layer to perform the prediction. Hence, this model is data-regularized due to the reconstruction cost and that makes it different from the CNN-LSTM model. We applied dropout after each LSTM layer as advised by \cite{zaremba2014recurrent}. Teacher forcing \citep{williams1989learning} was also applied to remove the reconstruction drift (bias) and increase the convergence speed. This model has a total of 32,027,456 parameters.

%%%%%%%%%%%%%%%%%%%%%%%%%%%%%%%%%%%%%%%%%%%%%%%%%%%%%%%%%%%%%%%%%%%%%%%%%%
\section{Results} 

%%%%%%%%%%%%%%%%%%%%%%%%%%%%%%%%%%%%%%%
\subsection{Evaluation Approach} 
% The prediction of diabetes onset is very important for the healthcare system as corresponding actions could be planned at the population level. 
% Specifically, the action could be based on the top $k$ percent of the population that is most likely to develop diabetes. 
%, or for all the targets classified as ones having diabetes using our model. As we output the \emph{probability} of getting diabetes in the future, we have to set a threshold for the second task. 
We use the area under receive-operation curve (AUC) as our primary metric for model evaluation. It is a standard metric for binary classification among related literature \cite{choi2019machine, leong2018prediction, sontag}. We also output the usual accuracy, sensitivity, specificity and positive predictive values for reference. In these cases, we just use the standard threshold of $0.5$.

%%%%%%%%%%%%%%%%%%%%%%%%%%%%%%%%%%%%%%%
\subsection{Model Performance}
\tableref{tab:result} shows the performance of all models. Note that we experimented with different values of window size $w$ and buffer size $b$ as well as the corresponding hyperparameters for training. Detailed results are listed in the supplementary materials. 

\tableResults

As seen, XGBoost has the best performance among all single models. Moreover, averaging the observations over all years seems to have a similar effect as concatenating them. This means that the temporal effect is not as useful in the XGBoost model. It is interesting to notice the disappointing performance of LSTM-based models, comparable to non-temporal model \emph{LR-avg} and \emph{Highway-avg} in this case. 

%\tableWindowSize

We also show the corresponding AUCs in \figureref{fig:result-buffer} with different observation window sizes $w$ for various models and different buffer sizes $b$ for our XGBoost model. Here, we used both concatenation and averaging aggregation over all years for different window sizes. It is evident that the larger the observation window, the more accurate the model would be. But, we observe that for $w=10$ in both the XGBoost and Highway models, the model performs slightly worse than $w=5$. This is an indicator that the observation data in the far past does not contribute much to the prediction performance, possibly due to adding more noise or mismatching data distributions. %Next, we look at the effect of the buffer size on our XGBoost-avg model with an observation window size $w=5$, as illustrated in Figure \ref{fig:result-buffer}. We exclude XGBoost-concat plot as the above results demonstrate that averaging the data gives similar results.  Finally, as expected,

We can also see that the larger the buffer size, the lower the performance of the XGBoost model, as the task becomes more challenging. It is interesting that even for a buffer of 10 years, the model performance is still better than the models with only fixed features (See \sectionref{section:ablation-study}). We see that AUC is maintained above 70 for up to six years ahead prediction. 

\figureXGBBufferY

%%%%%%%%%%%%%%%%%%%%%%%%%%%%%%%%%%%%%%%
\subsection{Feature Contribution}
For all models, we can compute how much each feature contributes to the results. We use the \emph{additive feature contribution framework}. Given a model $y = \sigma(f(X))$, and an input sample $\vec{x}\in\mathbb{R}^m$, where $\sigma$ is the logistic function and $X$ is the input, the framework will output the \emph{feature contribution} $\phi_j$ for feature $x_j$, which satisfies \emph{efficiency},
\[
\sum_j \phi_j = f(\vec{x}) - E_X(f(X)).
\]

As $f(x)$ is in the logit space, the feature contribution $\phi_j$ can be interpreted as the \emph{log odds ratio} between toggling the feature on and off.

Note that as the framework requires an input sample, we aggregate it using the sum of absolute values to get the feature contribution of each feature across all samples in the \emph{test set}. Absolute values are used so that positive and negative effects do not cancel each other across different samples.\footnote{Note that the aggregated feature contribution values do \emph{not} tell the correlation between the feature values with respect to the outcome.}

For LR and XGBoost, we used the Shapley value \citep{shapley, treeshap} as our feature contributions. For feed-forward neural network models, we used integrated gradients \citep{integrated-gradients} as our feature contributions method, which is an approximation of the Shapley value \citep{integrated-gradients}.

%In our study, we look at the Shapley Values (\cite{shapley}) for each feature. \red{WHAT ABOUT OTHER MODELS?!!! ===> This can be applied to both logistic regression and the XGBoost models}. Given a model $y = \sigma(f(X))$, and an input sample $x\in\mathbb{R}^m$, where $\sigma$ is the logistic function and $X$ is the input, the Shapley value $\phi_j$ for feature $x_j$ is defined as:
% \[
% \phi_j = \sum_{S\subseteq \{ x_1, \ldots, x_m\}\setminus \{x_j\}} \frac{|S|!(m - |S| - 1)!}{m!} (F(S\cup \{ x_j\}) - F(S)),
% \]
% where 
% \[
% F(S) = E_X(f(X) | X_S), \quad X_S := \{ x_i | i \in S\}.
% \]
% A main advantage of Shapley value is that it satisfies the \emph{efficiency},
% \[
% \sum_j x_j = f(x) - E_X(f(X)).
% \]

% Since the Shapley values satisfy the efficiency condition, and $f(x)$ is in the logit space, we can interpret the the Shapley value of each feature as the \emph{log odds ratio} between toggling the feature on and off.

% As Shapley values require an input sample, we aggregate it using the sum of absolute values to get the feature contribution of each feature. Absolute values are used so that positive and negative effects do not cancel each other across different samples.

% \paragraph{Logistic Regression}
% The computation of Shapley value is straight forward:
% \[
% \phi_j = \beta_j(x_j - E_X (X_j)),
% \]
% where $\beta_j$ is the coefficient of $X_j$. 

% \paragraph{XGBoost}
% The computation of Shapley value for XGBoost is provided in the XGBoost package using the TreeShap algorithm introduced by \cite{treeshap}.

\paragraph{Results}
We listed the top $10$ features for logistic regression in \tableref{tab:contrib-lr}. As seen, the fasting glucose and glucose in serum or plasma are contributing a lot, with an aggregated Shapley value of about $0.65$. We also observe that the geographical region contributes a lot to our prediction using logistic regression. It's interesting to note the appearance of landing date here, which takes non-null values only for immigrants as it's their date of arrival in Canada.

\tableAllFeatureContrib

The top 10 features for XGBoost are in \tableref{tab:contrib-xgb}. As seen, A1c contributes the most (i.e. an aggregated Shapley value of 0.908), followed by lab values. Prevalence of asthma in the top 10 corroborates studies stating that the use of steroids in the treatment of asthma can result in an increase in blood sugar level \citep{asthma-diabetes}. There is also correlation between individuals with diabetes having increased risk of getting asthma and other chronic diseases \citep{asthma}. We also see that an OHIP (insurance) claim for hypertension and immunity appear among the top features, suggesting that they may be relevant to diabetes onset.

\tableref{tab:contrib-highway} shows the top 10 features for the Highway model. It is mainly lab values but it includes a geographical feature (longitude). Note that in the Highway model, age seems to be more important than the other two models. We can see that the contribution values are much higher than those of the logistic regression and XGBoost, but this can be shown in the distribution of the model predictions. (See supplementary materials)

In conclusion, it seems that the glucose and A1c are very important lab results as they appear in all three models. Geographical feature (LHIN, longitude among others) in logistic regression and Highway network indicate that the location does affect the probability of having diabetes based on the data we have \citep{ices} .

%\tableXGBFeatureContrib

%%%%%%%%%%%%%%%%%%%%%%%%%%%%%%%%%%%%%%%
\subsection{Evaluation on Different Demographics}
As our models are trained on the whole training set, we are interested in discovering how they would perform on each demographic. We thus report performance and feature contributions for different demographics. When not specified, the model is XGBoost-avg with $w=5$ and $b=1$ as it is our best model with \emph{avg} and second-best overall. It is only slightly worse than \emph{concat} all the years which takes a much longer time and much more resources.

% \paragraph{Simulated Ontario Population} 
% As our training and testing cohort are almost balanced (about 53 to 47) for individuals with and without diabetes, it is not a good representation of the general public. As the percentage of Ontario population having diabetes was about 7\% in 2016 based on Statistics Canada\footnote{There are different figures that suggest it may be more, but we chose the lowest percentage because that would put more challenge in our models.} \citep{ontario}, we sampled from the test set a new test set with this ratio and evaluated our models in this unbalanced test set. \tableref{tab:unbalanced} shows the results. Note that XGBoost and Highway Network perform better, while logistic regression performs worse than on the original test set.

% \tableUnbalancedModels

\paragraph{Age, Gender and Country of Origin}
\tableref{tab:demographic} shows the model performances for the demographic groups gender, age and country of origin. As we can see, our results scale to most subcategories, especially when the sample size is high. It is worth noting the decrease in AUC for people younger than 40.

The feature contributions for each demographic is detailed in the supplementary materials. It is worth noticing that the prevalence of asthma is relatively more important for individuals younger than 50. As for country of origin, landing date is an important feature for people from China. This is in line with \cite{chinese}.

% \paragraph{Gender} 
%  Table \ref{tab:gender} outlines the top features for our XGBoost model. As seen, the AUC and top features are very similar to each other. It is interesting that hypertension claim code is within the top 5 features for males but not females \red{LAURA'S COMMENTS?!!}.

\tableAllDemographics
%\tableGenderModels

%\tableGenderXGBTopFeatures

% \paragraph{Age}
% Table \ref{tab:model-age} provides information on model evaluation for certain age ranges. Table \ref{tab:contrib-age} outlines the top features for our XGBoost model. Note that for each age group, A1c, max Range and blood A1c are always among the top 5 features. Age and birth year are important features when we restrict our model to young individuals and very old individuals. \red{Note that Shapley value aggregation does not indicate that whether the feature will have a positive or negative impact on the outcome}. It is also interesting that the prevalence of asthma is \emph{relatively} more important for people aged between $30$ to $49$.

%\tableAgeModels

%\tableAgeXGBTopFeatures

% \paragraph{Country of Origin} Table \ref{tab:model-country} evaluates our models on individuals born in a certain country. Table \ref{tab:contrib-country} outlines the top features for our XGBoost model. It is very interesting that landing date is a very important factor for people from China. Landing date is the date that the individual immigrates to Canada, and is only available for immigrants. \red{REMOVE? == > It is still unknown whether this is true or not, or it is just because of their age. - I thought landing date is just for immigrants only?}

%\tableCountryModels

%\tableCountryXGBTopFeatures

%%%%%%%%%%%%%%%%%%%%%%%%%%%%%%%%%%%%%%%
\subsection{Ablation Study}\label{section:ablation-study}
\paragraph{Datasets} We investigated the effect of each dataset in the performance of our model framework. \tableref{tab:pruning} outlines the results. It is clear that OHIP and OLIS contain a lot of effective features that frequently show up among the top features. It is also worth noting that combining all datasets still gives much better performance. We outline the most important features for each dataset in the corresponding table in the supplementary material.

\begin{table}[htbp]
\centering
\caption{Model performance when we only train on the mentioned datasets. We use XGBoost-avg here with $w = 5$ and $b = 1$}
\resizebox{\textwidth}{!}{
\begin{tabular}{cccccccccc}
Dataset(s) & All & Fixed & Chronic & DAD & ERCLAIM & NACRS & ODB & OHIP & OLIS \\
\hline
AUC & 79.9 & 62.6 & 63.1 & 56.7 & 57.9 & 58.6 & 59.4 & 69.8 & 72.6 \\ \hline
\end{tabular}
}
\label{tab:pruning}
 \end{table}
 
\paragraph{Top Features} Similarly, we perform pruning at the feature level by training the model only on the top $k$ features. \tableref{tab:pruning-data} summarizes the results. As seen, our top feature A1c only achieves an AUC of $68.8$. On the other hand, using the top $15$ features is approximately as good as using all features, suggesting the possibility of having a decent model by only using a small subset of features. Our study points out relevant features for further studies hand-picking input features.

\begin{table}[htbp]
\centering
\caption{Model performance when we only train on the top $k$ features. We use XGBoost-avg here with $w = 5$ and $b = 1$}
\begin{tabular}{cccccc}
Number of top features used & 1 & 5 & 10 & 15 & 963 (All) \\ \hline
AUC & 68.8 & 70.9 & 76.4 & 79.1 & 79.9\\ \hline
\end{tabular}
\label{tab:pruning-data}
\end{table}

%%%%%%%%%%%%%%%%%%%%%%%%%%%%%%%%%%%%%%%%%%%%%%%%%%%%%%%%%%%%%%%%%%%%%%%%%%
\section{Discussion} 

%[\red{LAURA'S FEEDBACK NEEDED}]Under construction
%%%%%%%%%%%%%%%%%%%%%%%%%%%%%%%%%%%%%%%
\subsection{Technical Significance - Revisited}

The examined models, i.e. XGBoost, regularized logistic regression, (Highway) multi layer perceptron, CNN-LSTM and LSTM Seq2Seq are among the most frequently used models in the recent literature. XGBoost performs the best among the examined models for diabetes prediction using an observation window of five years and a buffer of one year (i.e. predicting one year ahead). This is consistent with other studies using ICD codes \citep{nagata2018prediction}.

One of the challenges entangled with our cohort is the sparsity of features. For the raw data, only $44$ features out of $963$ are non-zero, when averaged over years and patients (see \tableref{tab:feature}). XGBoost has been shown to perform extremely well on tabular and sparse data \citep{xgboost}, particularly compared to deep-learning alternatives as demonstrated in \citep{nagata2018prediction} and our work.
%On the contrary, as mentioned in \citep{ruder2016overview}, when updating weights with stochastic gradient descent as often done with neural networks, it is better to perform a large update with a larger learning rate on rarely occurring features. In our study, we do not adapt neural network training for such data sparsity. 

LSTM-Seq2Seq with teaching forcing was found to outperform CNN-LSTM \citep{tonekaboni2018prediction}. This is expected as sequence to sequence models benefit from data-based regularization \citep{ghasedi2017deep}. However, both our temporal LSTMs do not improve on the Highway network. This is counter-intuitive as temporal models should aggregate important information over the input sequence, enriching the pre-classification states. But recurrent networks are hard to train \citep{pascanu2013difficulty}, and in our case, it is possible that the aggregation within each year constrained the learning capability of recurrent neural networks. %\red{DO YOU THINK CITING DEPICT IS A GOOD IDEA? THE POINT IS EXPLICITELY MENTIONED THERE.}.

%%%%%%%%%%%%%%%%%%%%%%%%%%%%%%%%%%%%%%%
\subsection{Clinical Relevance - Revisited} 
%-What features contributed the most and what they are
As observed, A1c and fasting glucose contribute the most to the prediction of diabetes. It confirms the common practice that A1c is used in diabetes tests to categorize individuals into having no diabetes, being pre-diabetic or living with diabetes \citep{a1c_test}. 

Our study also finds that the prevalence of asthma will affect the diabetes onset \citep{asthma-diabetes} in Ontario. This could lead to taking further steps for individuals with asthma on their prescribed steroids as it increases blood sugar, that might eventually result in weight gain and diabetes diagnosis \citep{asthma-diabetes}.

Our findings regarding hypertension contribution confirms the corresponding literature studies \citep{schutta2007diabetes, berraho2012hypertension, hashemizadeh2013hypertension}. Besides, geographical information plays a role too, recommending targeted interventions in higher risk areas.

As noted in \cite{chinese}, there is an increase in Chinese-Canadian diabetes incidence between 1996--2005. Landing date is one of the important features we study in predicting onset of diabetes for Chinese immigrants (or permanent residents). Hence, it is worth further research on to figure out the reason behind this fact. We also found that it is a harder task to predict diabetes on people under $40$ years old. It was found that approximately $80-92\%$ of patients with early onset of diabetes are obese compared to only $56\%$ of older ones \cite{wilmot2014early}. We do not have obesity nor BMI as input features, which probably makes it harder for our model to capture the specifics of early diabetes onset.

% \red{This might be due to the inverse linear relationship between BMI and diagnosis age of diabetes type 2 \cite{wilmot2014early}. It was found that approximately $80-92\%$ of patients with early onset of diabetes are obese compared to only $56\%$ of older ones \cite{wilmot2014early}}.
%\red{Why? less measurements (OLIS)? less interaction with OHIP? Less data in general for younger people?}

Our diabetes prediction method can be run without requiring additional data sources, as it relies on routine health data. It is modular, can be applied to any individual and generalizes well across all subcategories of gender, age and country of origin. If low computational resources are imposed, using $15$ features would still get a strong AUC. 

%%%%%%%%%%%%%%%%%%%%%%%%%%%%%%%%%%%%%%%
\subsection{Limits} 

The data that we use has numerous limitations. BMI has been shown to be a key component of diabetes pre-screening (\cite{edelstein1997predictors}) but is missing in our datasets. Our study lacks other important lab values such as cholesterol and glutamate decarboxylase antibodies. Finally, we only have country of origin as a proxy for ethnicity. All of this affects our model's prediction capacity. 

%%%%%%%%%%%%%%%%%%%%%%%%%%%%%%%%%%%%%%%%%%%%%%%%%%%%%%%%%%%%%%%%%%%%%%%%%%
\subsection{Future Work}

\paragraph{Predicting diabetes from self-reported results}
A similar (yet distinct) task compared to ours is predicting whether an individual will get diabetes in a range of $5-10$ years based on yearly survey data \citep{dport}. Surveys typically include some more personal level information such as dietary practice, physical activity, stress level and BMI, which are absent from the datasets used in this study. It is also possible to propose a model to make personalized recommendations for individuals at higher risk of diabetes. 

\paragraph{Diabetes clustering}
Diabetes can be roughly classified into 3 different types (Type I, Type II and gestational diabetes). It is believed that type II is very heterogeneous, and recent work by \cite{sweden} applied a clustering method using six important features to cluster a  diabetes cohort into five types. We plan to explore different clustering methods (including the deep learning ones) on our diabetes cohort. A potential clinical application is that different types can be treated differently, i.e. a major step towards personalized treatment. 

%%%%%%%%%%%%%%%%%%%%%%%%%%%%%%%%%%%%%%%%%%%%%%%%%%%%%%%%%%%%%%%%%%%%%%%%%%
\section{Conclusion}
In this study, we demonstrated the effectiveness of machine learning algorithms to predict future onset of diabetes considering an individual's medical history over the last few years. Using a single XGBoost model, we reached a test-AUC of $80.3$, predicting diabetes one year ahead, which outperforms any other model we examined. The provided features' contribution show that lab results (led by A1c) are essential for the task, which advocates for A1c screening. Other chronic diseases' flags (such as asthma, hypertension, etc) as well as diagnosis codes in insurance claims are among the top feature contributors, containing signals on future diabetes incidence more than five years ahead.

% ACKNOWLEDGEMENTS ONLY GO IN THE CAMERA-READY, NOT THE SUBMISSION
% \acks{Many thanks to all collaborators and funders!}

\bibliography{refs}

\begin{thebibliography}{47}
\providecommand{\natexlab}[1]{#1}
\providecommand{\url}[1]{\texttt{#1}}
\expandafter\ifx\csname urlstyle\endcsname\relax
  \providecommand{\doi}[1]{doi: #1}\else
  \providecommand{\doi}{doi: \begingroup \urlstyle{rm}\Url}\fi

\bibitem[Ahlqvist et~al.(2018)Ahlqvist, Storm, Käräjämäki, Martinell,
  Dorkhan, Carlsson, Vikman, Prasad, Aly, Almgren, Wessman, Shaat, Spégel,
  Mulder, Lindholm, Melander, Hansson, Malmqvist, Åke Lernmark, Lahti,
  Forsén, Tuomi, Rosengren, and Groop]{sweden}
Emma Ahlqvist, Petter Storm, Annemari Käräjämäki, Mats Martinell, Mozhgan
  Dorkhan, Annelie Carlsson, Petter Vikman, Rashmi~B Prasad, Dina~Mansour Aly,
  Peter Almgren, Ylva Wessman, Nael Shaat, Peter Spégel, Hindrik Mulder, Eero
  Lindholm, Olle Melander, Ola Hansson, Ulf Malmqvist, Åke Lernmark, Kaj
  Lahti, Tom Forsén, Tiinamaija Tuomi, Anders~H Rosengren, and Leif Groop.
\newblock Novel subgroups of adult-onset diabetes and their association with
  outcomes: a data-driven cluster analysis of six variables.
\newblock \emph{The Lancet Diabetes \& Endocrinology}, 6\penalty0 (5):\penalty0
  361 -- 369, 2018.
\newblock ISSN 2213-8587.
\newblock \doi{https://doi.org/10.1016/S2213-8587(18)30051-2}.
\newblock URL
  \url{http://www.sciencedirect.com/science/article/pii/S2213858718300512}.

\bibitem[Alangh et~al.(2013)Alangh, Chiu, and Shah]{chinese}
Avreet Alangh, Maria Chiu, and Baiju~R. Shah.
\newblock Rapid increase in diabetes incidence among chinese canadians between
  1996 and 2005.
\newblock \emph{Diabetes Care}, 2013.
\newblock ISSN 0149-5992.
\newblock \doi{10.2337/dc13-0052}.
\newblock URL
  \url{http://care.diabetesjournals.org/content/early/2013/05/28/dc13-0052}.

\bibitem[Alhassan et~al.(2018)Alhassan, McGough, Alshammari, Daghstani, Budgen,
  and Al~Moubayed]{alhassan2018type}
Zakhriya Alhassan, A~Stephen McGough, Riyad Alshammari, Tahani Daghstani, David
  Budgen, and Noura Al~Moubayed.
\newblock Type-2 diabetes mellitus diagnosis from time series clinical data
  using deep learning models.
\newblock In \emph{International Conference on Artificial Neural Networks},
  pages 468--478. Springer, 2018.

\bibitem[Atkinson et~al.(2014)Atkinson, Eisenbarth, and
  Michels]{atkinson2014type}
Mark~A Atkinson, George~S Eisenbarth, and Aaron~W Michels.
\newblock Type 1 diabetes.
\newblock \emph{The Lancet}, 383\penalty0 (9911):\penalty0 69--82, 2014.

\bibitem[Berraho et~al.(2012)Berraho, El~Achhab, Benslimane, Rhazi, Chikri, and
  Nejjari]{berraho2012hypertension}
Mohamed Berraho, Youness El~Achhab, Abdelilah Benslimane, K~EL Rhazi, Mohamed
  Chikri, and Chakib Nejjari.
\newblock Hypertension and type 2 diabetes: a cross-sectional study in morocco
  (epidiam study).
\newblock \emph{Pan African Medical Journal}, 11\penalty0 (1), 2012.

\bibitem[Bhardwaj et~al.(2018)Bhardwaj, Wodajo, Spano, Neal, and
  Coustasse]{bhardwaj2018impact}
Niharika Bhardwaj, Bezawit Wodajo, Anthony Spano, Symaron Neal, and Alberto
  Coustasse.
\newblock The impact of big data on chronic disease management.
\newblock \emph{The health care manager}, 37\penalty0 (1):\penalty0 90--98,
  2018.

\bibitem[Booth et~al.(2012)Booth, Polsky, Gozdyra, Cauch-Dudek, Kiran, Shah,
  Lipscombe, and H.]{ices}
G.~L. Booth, J.~Y. Polsky, P.~Gozdyra, K.~Cauch-Dudek, T.~Kiran, B.R. Shah,
  L.~L. Lipscombe, and Glazier~R. H.
\newblock Regional measures of diabetes burden in ontario. toronto.
\newblock 2012.

\bibitem[Chen and Guestrin(2016)]{xgboost}
Tianqi Chen and Carlos Guestrin.
\newblock Xgboost: A scalable tree boosting system.
\newblock In \emph{Proceedings of the 22Nd ACM SIGKDD International Conference
  on Knowledge Discovery and Data Mining}, KDD '16, pages 785--794, New York,
  NY, USA, 2016. ACM.
\newblock ISBN 978-1-4503-4232-2.
\newblock \doi{10.1145/2939672.2939785}.
\newblock URL \url{http://doi.acm.org/10.1145/2939672.2939785}.

\bibitem[Choi et~al.(2019)Choi, Rha, Kim, Kang, Park, and Noh]{choi2019machine}
Byoung~Geol Choi, Seung-Woon Rha, Suhng~Wook Kim, Jun~Hyuk Kang, Ji~Young Park,
  and Yung-Kyun Noh.
\newblock Machine learning for the prediction of new-onset diabetes mellitus
  during 5-year follow-up in non-diabetic patients with cardiovascular risks.
\newblock \emph{Yonsei medical journal}, 60\penalty0 (2):\penalty0 191--199,
  2019.

\bibitem[Ding et~al.(2018)Ding, Simpson, Pfohl, Kale, Jung, and
  Shah]{ding2018effectiveness}
Daisy~Yi Ding, Chlo{\'e} Simpson, Stephen Pfohl, Dave~C Kale, Kenneth Jung, and
  Nigam~H Shah.
\newblock The effectiveness of multitask learning for phenotyping with
  electronic health records data.
\newblock \emph{arXiv preprint arXiv:1808.03331}, 2018.

\bibitem[E~Hux et~al.(2002)E~Hux, Ivis, Flintoft, and Bica]{hux}
Janet E~Hux, Frank Ivis, Virginia Flintoft, and Adina Bica.
\newblock Diabetes in ontario: Determination of prevalence and incidence using
  a validated administrative data algorithm.
\newblock \emph{Diabetes care}, 25:\penalty0 512--6, 04 2002.
\newblock \doi{10.2337/diacare.25.3.512}.

\bibitem[Edelstein et~al.(1997)Edelstein, Knowler, Bain, Andres,
  Barrett-Connor, Dowse, Haffher, Pettitt, Sorkin, Muller,
  et~al.]{edelstein1997predictors}
Sharon~L Edelstein, William~C Knowler, Raymond~P Bain, Reubin Andres,
  Elizabeth~L Barrett-Connor, Gary~K Dowse, Steven~M Haffher, David~J Pettitt,
  John~D Sorkin, Denis~C Muller, et~al.
\newblock Predictors of progression from impaired glucose tolerance to niddm:
  an analysis of six prospective studies.
\newblock \emph{Diabetes}, 46\penalty0 (4):\penalty0 701--710, 1997.

\bibitem[Ehrlich et~al.(2010)Ehrlich, Quesenberry, Van Den~Eeden, Shan, and
  Ferrara]{asthma}
Samantha~F. Ehrlich, Charles~P. Quesenberry, Stephen~K. Van Den~Eeden, Jun
  Shan, and Assiamira Ferrara.
\newblock Patients diagnosed with diabetes are at increased risk for asthma,
  chronic obstructive pulmonary disease, pulmonary fibrosis, and pneumonia but
  not lung cancer.
\newblock \emph{Diabetes Care}, 33\penalty0 (1):\penalty0 55--60, 2010.
\newblock ISSN 0149-5992.
\newblock \doi{10.2337/dc09-0880}.
\newblock URL \url{http://care.diabetesjournals.org/content/33/1/55}.

\bibitem[Garske(2018)]{garske2018using}
Thomas Garske.
\newblock \emph{Using Deep Learning on EHR Data to Predict Diabetes}.
\newblock PhD thesis, University of Colorado at Denver, 2018.

\bibitem[Ghasedi~Dizaji et~al.(2017)Ghasedi~Dizaji, Herandi, Deng, Cai, and
  Huang]{ghasedi2017deep}
Kamran Ghasedi~Dizaji, Amirhossein Herandi, Cheng Deng, Weidong Cai, and Heng
  Huang.
\newblock Deep clustering via joint convolutional autoencoder embedding and
  relative entropy minimization.
\newblock In \emph{Proceedings of the IEEE International Conference on Computer
  Vision}, pages 5736--5745, 2017.

\bibitem[Harjutsalo et~al.(2008)Harjutsalo, Sj{\"o}berg, and
  Tuomilehto]{harjutsalo2008time}
Valma Harjutsalo, Lena Sj{\"o}berg, and Jaakko Tuomilehto.
\newblock Time trends in the incidence of type 1 diabetes in finnish children:
  a cohort study.
\newblock \emph{The Lancet}, 371\penalty0 (9626):\penalty0 1777--1782, 2008.

\bibitem[Hashemizadeh et~al.(2013)Hashemizadeh, Sarvelayati,
  et~al.]{hashemizadeh2013hypertension}
Haydeh Hashemizadeh, Dorari Sarvelayati, et~al.
\newblock Hypertension and type 2 diabetes: A cross-sectional study in
  hospitalized patients in quchan, iran.
\newblock \emph{Iranian Journal of Diabetes and Obesity}, 5\penalty0
  (1):\penalty0 21--26, 2013.

\bibitem[Ioffe and Szegedy(2015)]{ioffe2015batch}
Sergey Ioffe and Christian Szegedy.
\newblock Batch normalization: Accelerating deep network training by reducing
  internal covariate shift.
\newblock \emph{arXiv preprint arXiv:1502.03167}, 2015.

\bibitem[{Joyce Lee}(2018)]{joyce2018}
{Joyce Lee}.
\newblock {MD, MPH, Robert Kelch Professor of Pediatrics}.
\newblock \url{https://www.mlforhc.org/2018}, 2018.
\newblock Online; accessed 19 March 2019.

\bibitem[Krishnan et~al.(2013)Krishnan, Razavian, Choi, Nigam, Blecker,
  Schmidt, and Sontag]{krishnan2013early}
RahulG Krishnan, Narges Razavian, Youngduck Choi, Somesh Nigam, Saul Blecker,
  A~Schmidt, and David Sontag.
\newblock Early detection of diabetes from health claims.
\newblock In \emph{Machine Learning in Healthcare Workshop, NIPS}, 2013.

\bibitem[Leong et~al.(2018)Leong, Daya, Porneala, Devlin, Shiffman, McPhaul,
  Selvin, and Meigs]{leong2018prediction}
Aaron Leong, Natalie Daya, Bianca Porneala, James~J Devlin, Dov Shiffman,
  Michael~J McPhaul, Elizabeth Selvin, and James~B Meigs.
\newblock Prediction of type 2 diabetes by hemoglobin a1c in two
  community-based cohorts.
\newblock \emph{Diabetes care}, 41\penalty0 (1):\penalty0 60--68, 2018.

\bibitem[{Lundberg} et~al.(2018){Lundberg}, {Erion}, and {Lee}]{treeshap}
Scott~M. {Lundberg}, Gabriel~G. {Erion}, and Su-In {Lee}.
\newblock {Consistent Individualized Feature Attribution for Tree Ensembles}.
\newblock \emph{arXiv e-prints}, art. arXiv:1802.03888, Feb 2018.

\bibitem[Mikalsen(2019)]{mikalsen2019advancing}
Karl~{\O}yvind Mikalsen.
\newblock Advancing unsupervised and weakly supervised learning with emphasis
  on data-driven healthcare.
\newblock 2019.

\bibitem[Miotto et~al.(2016)Miotto, Li, and Dudley]{miotto2016deep}
Riccardo Miotto, Li~Li, and Joel~T Dudley.
\newblock Deep learning to predict patient future diseases from the electronic
  health records.
\newblock In \emph{European Conference on Information Retrieval}, pages
  768--774. Springer, 2016.

\bibitem[Nagata et~al.(2018)Nagata, Takai, Yasuda, Heracleous, and
  Yoneyama]{nagata2018prediction}
Masatoshi Nagata, Kohichi Takai, Keiji Yasuda, Panikos Heracleous, and Akio
  Yoneyama.
\newblock Prediction models for risk of type-2 diabetes using health claims.
\newblock In \emph{Proceedings of the BioNLP 2018 workshop}, pages 172--176,
  2018.

\bibitem[NIDDK(2018)]{a1c_test}
NIDDK.
\newblock How is the a1c test used to diagnose type 2 diabetes and prediabetes?
\newblock
  \url{https://www.niddk.nih.gov/health-information/diabetes/overview/tests-diagnosis/a1c-test},
  2018.
\newblock Accessed: 2019-03-28.

\bibitem[Osmani et~al.(2018)Osmani, Li, Danieletto, Glicksberg, Dudley, and
  Mayora]{osmani2018processing}
Venet Osmani, Li~Li, Matteo Danieletto, Benjamin Glicksberg, Joel Dudley, and
  Oscar Mayora.
\newblock Processing of electronic health records using deep learning: A
  review.
\newblock \emph{arXiv preprint arXiv:1804.01758}, 2018.

\bibitem[Pascanu et~al.(2013)Pascanu, Mikolov, and
  Bengio]{pascanu2013difficulty}
Razvan Pascanu, Tomas Mikolov, and Yoshua Bengio.
\newblock On the difficulty of training recurrent neural networks.
\newblock In \emph{International conference on machine learning}, pages
  1310--1318, 2013.

\bibitem[Razavian et~al.(2015{\natexlab{a}})Razavian, Blecker, Schmidt,
  Smith-McLallen, Nigam, and Sontag]{razavian2015population}
Narges Razavian, Saul Blecker, Ann~Marie Schmidt, Aaron Smith-McLallen, Somesh
  Nigam, and David Sontag.
\newblock Population-level prediction of type 2 diabetes from claims data and
  analysis of risk factors.
\newblock \emph{Big Data}, 3\penalty0 (4):\penalty0 277--287,
  2015{\natexlab{a}}.

\bibitem[Razavian et~al.(2015{\natexlab{b}})Razavian, Blecker, Schmidt,
  Smith-McLallen, Nigam, and Sontag]{sontag}
Narges Razavian, Saul Blecker, Ann~Marie Schmidt, Aaron Smith-McLallen, Somesh
  Nigam, and David Sontag.
\newblock Population-level prediction of type 2 diabetes from claims data and
  analysis of risk factors.
\newblock \emph{Big Data}, 3\penalty0 (4):\penalty0 277--287,
  2015{\natexlab{b}}.

\bibitem[Rosella et~al.(2011)Rosella, Manuel, Burchill, Stukel, and ]{dport}
Laura~C Rosella, Douglas~G Manuel, Charles Burchill, Th{\'e}r{\`e}se~A Stukel,
  and .
\newblock A population-based risk algorithm for the development of diabetes:
  development and validation of the diabetes population risk tool (dport).
\newblock \emph{Journal of Epidemiology \& Community Health}, 65\penalty0
  (7):\penalty0 613--620, 2011.
\newblock ISSN 0143-005X.
\newblock \doi{10.1136/jech.2009.102244}.
\newblock URL \url{https://jech.bmj.com/content/65/7/613}.

\bibitem[Rosella et~al.(2012)Rosella, Mustard, Stukel, Corey, Hux, Roos, and
  Manuel]{rosella2012role}
Laura~C Rosella, Cameron~A Mustard, Therese~A Stukel, Paul Corey, Jan Hux, Les
  Roos, and Douglas~G Manuel.
\newblock The role of ethnicity in predicting diabetes risk at the population
  level.
\newblock \emph{Ethnicity \& health}, 17\penalty0 (4):\penalty0 419--437, 2012.

\bibitem[Rosella et~al.(2016)Rosella, Lebenbaum, Fitzpatrick, O'reilly, Wang,
  Booth, Stukel, and Wodchis]{rosella2016impact}
LC~Rosella, M~Lebenbaum, T~Fitzpatrick, D~O'reilly, J~Wang, GL~Booth,
  TA~Stukel, and WP~Wodchis.
\newblock Impact of diabetes on healthcare costs in a population-based cohort:
  a cost analysis.
\newblock \emph{Diabetic Medicine}, 33\penalty0 (3):\penalty0 395--403, 2016.

\bibitem[Samer El~Jerjawi and Abu-Naser(2018)]{eljerjawi}
Nesreen Samer El~Jerjawi and Samy Abu-Naser.
\newblock Diabetes prediction using artificial neural network.
\newblock \emph{Journal of Advanced Science}, 124:\penalty0 1--10, 12 2018.

\bibitem[Schutta(2007)]{schutta2007diabetes}
Mark~H Schutta.
\newblock Diabetes and hypertension: epidemiology of the relationship and
  pathophysiology of factors associated with these comorbid conditions.
\newblock \emph{Journal of the cardiometabolic syndrome}, 2\penalty0
  (2):\penalty0 124--130, 2007.

\bibitem[Shapley(1953)]{shapley}
Lloyd~S Shapley.
\newblock A value for n-person games.
\newblock In Harold~W. Kuhn and Albert~W. Tucker, editors, \emph{Contributions
  to the Theory of Games II}, pages 307--317. Princeton University Press,
  Princeton, 1953.

\bibitem[Srivastava et~al.(2015)Srivastava, Greff, and Schmidhuber]{highway}
Rupesh~Kumar Srivastava, Klaus Greff, and J{\"u}rgen Schmidhuber.
\newblock Highway networks.
\newblock \emph{CoRR}, abs/1505.00387, 2015.

\bibitem[Suissa et~al.(2010)Suissa, Kezouh, and Ernst]{asthma-diabetes}
Samy Suissa, Abbas Kezouh, and Pierre Ernst.
\newblock Inhaled corticosteroids and the risks of diabetes onset and
  progression.
\newblock \emph{The American Journal of Medicine}, 123\penalty0 (11):\penalty0
  1001--1006, Nov 2010.
\newblock ISSN 0002-9343.
\newblock \doi{10.1016/j.amjmed.2010.06.019}.
\newblock URL \url{https://doi.org/10.1016/j.amjmed.2010.06.019}.

\bibitem[Sundararajan et~al.(2017)Sundararajan, Taly, and
  Yan]{integrated-gradients}
Mukund Sundararajan, Ankur Taly, and Qiqi Yan.
\newblock Axiomatic attribution for deep networks.
\newblock \emph{CoRR}, abs/1703.01365, 2017.
\newblock URL \url{http://arxiv.org/abs/1703.01365}.

\bibitem[Sutskever et~al.(2014)Sutskever, Vinyals, and
  Le]{sutskever2014sequence}
Ilya Sutskever, Oriol Vinyals, and Quoc~V Le.
\newblock Sequence to sequence learning with neural networks.
\newblock In \emph{Advances in neural information processing systems}, pages
  3104--3112, 2014.

\bibitem[Thesmar et~al.(2019)Thesmar, Sraer, Pinheiro, Dadson, Veliche, and
  Greenberg]{thesmar2019combining}
David Thesmar, David Sraer, Lisa Pinheiro, Nick Dadson, Razvan Veliche, and
  Paul Greenberg.
\newblock Combining the power of artificial intelligence with the richness of
  healthcare claims data: Opportunities and challenges.
\newblock \emph{PharmacoEconomics}, pages 1--8, 2019.

\bibitem[Tonekaboni et~al.(2018)Tonekaboni, Mazwi, Laussen, Eytan, Greer,
  Goodfellow, Goodwin, Brudno, and Goldenberg]{tonekaboni2018prediction}
Sana Tonekaboni, Mjaye Mazwi, Peter Laussen, Danny Eytan, Robert Greer,
  Sebastian~D Goodfellow, Andrew Goodwin, Michael Brudno, and Anna Goldenberg.
\newblock Prediction of cardiac arrest from physiological signals in the
  pediatric icu.
\newblock In \emph{Machine Learning for Healthcare Conference}, pages 534--550,
  2018.

\bibitem[Williams and Zipser(1989)]{williams1989learning}
Ronald~J Williams and David Zipser.
\newblock A learning algorithm for continually running fully recurrent neural
  networks.
\newblock \emph{Neural computation}, 1\penalty0 (2):\penalty0 270--280, 1989.

\bibitem[Wilmot and Idris(2014)]{wilmot2014early}
Emma Wilmot and Iskandar Idris.
\newblock Early onset type 2 diabetes: risk factors, clinical impact and
  management.
\newblock \emph{Therapeutic advances in chronic disease}, 5\penalty0
  (6):\penalty0 234--244, 2014.

\bibitem[Wu et~al.(2018)Wu, Yang, Huang, He, and Wang]{wu2018type}
Han Wu, Shengqi Yang, Zhangqin Huang, Jian He, and Xiaoyi Wang.
\newblock Type 2 diabetes mellitus prediction model based on data mining.
\newblock \emph{Informatics in Medicine Unlocked}, 10:\penalty0 100--107, 2018.

\bibitem[Zaremba et~al.(2014)Zaremba, Sutskever, and
  Vinyals]{zaremba2014recurrent}
Wojciech Zaremba, Ilya Sutskever, and Oriol Vinyals.
\newblock Recurrent neural network regularization.
\newblock \emph{arXiv preprint arXiv:1409.2329}, 2014.

\bibitem[Zou et~al.(2018)Zou, Qu, Luo, Yin, Ju, and Tang]{zou2018predicting}
Quan Zou, Kaiyang Qu, Yamei Luo, Dehui Yin, Ying Ju, and Hua Tang.
\newblock Predicting diabetes mellitus with machine learning techniques.
\newblock \emph{Frontiers in genetics}, 9, 2018.

\end{thebibliography}

% \appendix
% \section*{Appendix A.}

% \tableICDNineCodeMap

% \tableICDTenCodeMap

% \tableLRParameters

% \tableXGBParameters

% \figureHighwayArchitecture

% \tableXGBAvgTopFeatures

% \figureXGBBalanced

% \figureXGBUnbalanced

% \figureLRBalanced

% \figureLRUnbalanced

% \figureHighwayBalanced

% \figureHighwayUnbalanced

% \tableDatasetPruning

% \tableDatasetCoverage

\end{document}